\documentclass[aps,prl,groupedaddress,nofootinbib,reprint]{revtex4-2}
\usepackage{graphicx}

\newcommand{\ud}{\mathrm{d}}
\newcommand{\um}{\mu \mathrm{m}}
\newcommand{\nm}{\mathrm{nm}}
\newcommand{\eps}{\varepsilon}

\newcommand{\MHz}{\mathrm{MHz}}
\newcommand{\picom}{\mathrm{pm}}
\usepackage{amsmath}

\usepackage{physics}
\renewcommand{\div}{\divisionsymbol}
\usepackage{cancel}

\begin{document}
\title{Gigahertz modulation of a fully dielectric nonlocal metasurface}

\author{Alessandro Pitanti}
\affiliation{CNR-Istituto Nanoscienze, Laboratorio NEST, Piazza San Silvestro 12, 56127 Pisa, Italy}
\affiliation{Paul-Drude-Institut für Festkörperelektronik, Leibniz-Institut im Forschungsverbund Berlin e.~V., 5-7, Hausvogteiplatz, Berlin, 10117 (Germany)}

\author{Gaia Da Prato}
\affiliation{Dipartimento di Fisica “E. Fermi”, Università di Pisa, Largo Pontecorvo 3, 56127 Pisa (Italy)}
\altaffiliation{Present address: Kavli Institute of Nanoscience, Department of Quantum Nanoscience, Delft University of Technology, 2628CJ Delft, The Netherlands}

\author{Giorgio Biasiol}
\affiliation{CNR-Istituto Officina dei Materiali, Laboratorio TASC, 34149 Trieste (Italy)}

\author{Alessandro Tredicucci}
\affiliation{Dipartimento di Fisica “E. Fermi”, Università di Pisa and CNR-Istituto Nanoscienze, Largo Pontecorvo 3, 56127 Pisa (Italy)}

\author{Simone Zanotto}
\affiliation{CNR-Istituto Nanoscienze, Laboratorio NEST, Piazza San Silvestro 12, 56127 Pisa, Italy}
\email{simone.zanotto@nano.cnr.it}

\begin{abstract}
Nonlocal metasurfaces are currently emerging as advanced tools for the manipulation of electromagnetic radiation, going beyond the widely explored Huygens metasurface concept. Nonetheless, the lack of an unified approach for their fast and efficient tunability still represents a serious challenge to overcome. In this article we report on gigahertz modulation of a dielectric slab-based, nonlocal (i.e.~angle-dispersive) metasurface, whose operation relies on the optomechanical coupling with a mechanical wave excited piezoelectrically by a transducer integrated on the same chip. Importantly, the metasurface region is free from any conductive material, thus eliminating optical losses, and making our device of potential interest for delicate environments such as high-power apparatuses or quantum optical systems.
\end{abstract}

\maketitle

\section{Introduction}
In the last decade, the advent of metasurfaces has brought a revolution in optics, leading to ultracompact lenses, polarization elements, and other special beam manipulation components \cite{Chen_2016, Hu2021}. The great majority of those objects are however developed within a quite special framework: that of local response. In this approach, each metasurface element (meta-atom) emits radiation that depends only on the incident field exciting that precise meta-atom. While this concept eases the design of metasurfaces - that can be basically constructed as phase, amplitude, or polarization holograms assembling pre-designed unit cells - it possesses intrinsic limitations that can only be overcome by releasing the assumption of local response. In a nonlocal metasurface, the radiation from a meta-atom also depends on the field that excites neighboring elements; equivalently, the response of the metasurface to plane waves is dependent on the angle of incidence \cite{Shastri2023}. It has been shown that nonlocal metasurfaces can perform complex operations such as optical signal processing \cite{Kwon2018}, image differentiation and edge detection \cite{Guo2018}, phase contrast imaging \cite{Ji2022}, perfect anomalous reflection \cite{He2022}, multifunctional wavefront shaping \cite{Malek2022}, and space compression \cite{Shastri2022,Miller2023}. 

In addition to relieving the limitations of local response, metasurface scientists have been focusing on two other challenging fronts: optical losses mitigation and reconfigurability/modulation. To reduce losses, great efforts have been devoted to the development of dielectric metasurfaces, that are now close to technological maturity \cite{Kamali2018, Hu2020, Huang2023}. Meanwhile, the need for reconfiguration and modulation has been constantly increasing: several mechanisms and materials have been reported to dynamically tune the response of meta-atoms, with speed, energy consumption, ease of fabrication, and electrical control as main figures of merit to be considered \cite{Badloe2021,Yang2022,Guanxing2021,Jung2021}. 

Despite the efforts to advance over all the aforementioned fronts, it appears that to date no study has addressed simultaneously all the three issues of (i) nonlocal response, (ii) absence of optical losses, and (iii) electrically driven high-frequency modulation, greatly limiting the potentials of applying metasurfaces technologies to high-power optics, quantum photonics and light-based communications. For instance, multi-gigahertz (GHz) metasurface actuation has been observed, without however direct electrical driving \cite{Ajia2021, OBrien2014} . Electric control of metasurfaces has already been reported in the last decade \cite{Ou2013,Zhang2020} and is still the subject of recent advances \cite{Benea-Chelmus2022}; however, it required conducting elements extending into the optically active metasurface region, inducing optical losses and consequent heating that potentially hinders applications in high-power laser systems. Other reports investigated electrically driven, fully dielectric, and potentially nonlocal metasurfaces; however, the response was there limited to DC or few tens kHz \cite{Kwon2021, Kwon2022, Li2019, Mansha2022}. 

To overcome those limitations, we have conceived an optomechanical dielectric-slab-based device, where GHz vibrations excited by an interdigitated transducer via the piezoelectric effect propagate towards the photonic-phononic crystal that constitutes the modulating metasurface \cite{Hussein2014}. Metallic IDTs are spatially separated from the optically active region, thus completely eliminating optical losses. We developed our device on a monolithic gallium arsenide (GaAs) based platform, extending previous knowledge \cite{Lima2005, Courjal2010, Kim2001, Tadesse2014} to a membrane-based, free-standing metasurface, which allows for high permittivity contrast and high-quality factor (Q) dispersive resonances \cite{Zanotto2019}. Moreover, the treatment of photoelastic and moving boundary response of GaAs-based nanoresonators is well established, thus easing the setting of a theoretical background \cite{Baker2014, Balram2014}. GaAs also compares favorably with other platforms as far as certain metrics are considered. For instance, while aluminum nitride (AlN) has clear advantages regarding integration with silicon \cite{Dong2022, Luo2023}, AlN cannot host quantum dot and quantum well emitters and detectors, that can be instead realized in GaAs \cite{Fuhrmann2011, Wigger2021}, with applications spanning from single photon manipulation \cite{Buhler2022} to terahertz technology. In comparison with lithium niobate (LiNbO3), GaAs has weaker piezoelectric coefficients \cite{Jiang2019, Hassanien2021, Klopfer2022, Marinkovic2021, Wan2022, Weigand2021, Weiss2022}; however, its fabrication processes is more standard and less demanding.

\section{Results}
Our study relies on the device represented in Fig.~1a-c, namely, a GaAs/AlGaAs-based monolithic heterostructured chip, where interdigitated transducers (IDTs) are placed in close proximity of a substrate-free region where the metasurface is defined. 
\begin{figure*}
\centering
\includegraphics[width = \textwidth]{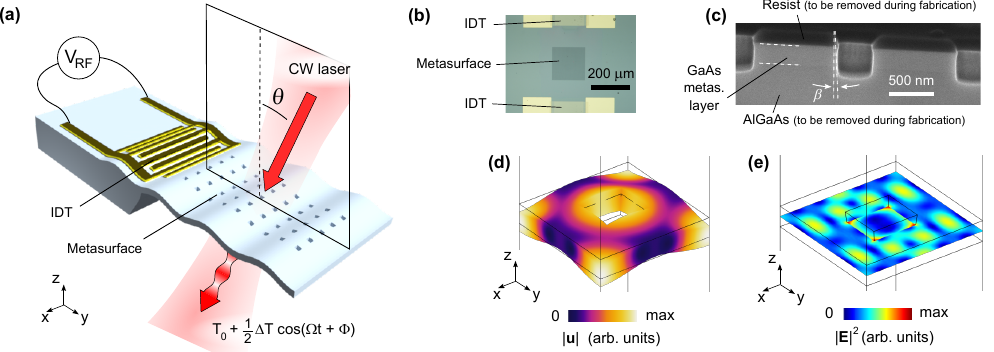}
\caption{Device concept and resonant modes. (a), Rendered scheme of optomechanical modulation in a nonlocal metasurface. A radiofrequency (RF) voltage is applied to an interdigitated electrode (IDT) monolithically integrated on a piezoelectric chip. Mechanical waves propagate into the metasurface region (consisting of a hole array on a membrane) where coupling with the optical angle-dispersive resonance occurs. The effect is observed as a modulation of transmitted light intensity. A microscope picture of a portion of the chip is reported in (b). Panel (c) depicts a scanning electron microscope image of a sacrificial chip cross-section collected at an intermediate fabrication step, highlighting the hole sidewall shape. (d) and (e) illustrate, respectively, the simulated mechanical displacement field and the optical electric field of the main Bloch eigenmodes involved in the optomechanical coupling process.}
\end{figure*}
Fabrication details are given in Supplementary Sect. I. The metasurface consists of a hole array patterned on a quasi-square lattice in the dielectric membrane (pitches $a_x = 1324\ \nm$, $a_y = 1317\ \nm$). The holes have an almost square shape in the $x-y$ plane ($x$ and $y$ sides lengths of 450 nm and 395 nm, respectively), and a slightly slanted profile in the vertical plane ($\beta = 5^{\circ}$), as revealed by the cross section of Fig.~1c (collected from a sacrificial sample at an intermediate processing step). The chip is wire-bonded to a printed circuit board (PCB) and mounted on a rotation-translation stage. Thanks to the intrinsic piezoelectricity of GaAs, the IDTs excite a surface acoustic wave (SAW) propagating towards the metasurface membrane; eventually, in the patterned region, the excitation consists of mechanical Bloch modes. In our device, mechanical frequencies in the range 960-1120 MHz can be excited, thanks to IDTs with different pitch (periods from 2.5 to $2.85\ \um$) facing distinct, but geometrically identical, metasurfaces. Figure 1d represents the simulated profile of the mechanical Bloch mode that plays the most important optomechanical role in our sample. The sample is illuminated with a weakly focused ($30\ \um$) $x$-polarized laser beam, impinging on the metasurface at an angle $\theta$ (further details on the setup in Suppl.~Sect.~II). When the beam wavelength-angle pair matches the dispersive resonance condition, a photonic resonance is excited; a detailed analysis of the resonance dispersion is provided in Suppl.~Sect.~III. In Fig.~1e we report the simulated local field for the specific resonance analyzed in this article. 

The mechanical Bloch modes within the metasurface perturbs both the hole shape and refractive index, generating a modulation effect on the photonic resonances coupled with them. To develop a quantitative model of the modulation effect, we relied on the quasi-normal mode perturbation theory (QNMPT), elaborating on the concepts and methods illustrated in \cite{Primo2020, Yan2020, Zhao2019, Liu2022}. Quasi-normal modes are eigenstates, corresponding to complex eigenvalues, which emerge from radiative and non-radiative energy losses of open resonators. In our model, another key assumption is an approximation analogous to the Born-Oppenheimer one: since the time scales of near-infrared light waves and that of mechanical vibrations differ by orders of magnitude, we calculated the electromagnetic (e.m.) response assuming that the metasurface shape is “frozen” at a given time within the mechanical cycle. Calculating the e.m. response for all the times of the mechanical cycle leads to the time-dependent (i.e., modulated) e.m. response of the optomechanical metasurface. Moreover, since the shape and refractive index changes induced by the mechanical displacement field are small, one is allowed to adopt a perturbative approach over a suitable basis of electromagnetic Bloch quasi-normal modes. In our case, especially when quasi-normal incidence is considered, the photonic resonance is close to being twofold degenerate, thus calling in principle for a two-mode perturbation theory. However, it can be shown (see Suppl.~Sect.~IV) that, as long as the photonic and the mechanical Bloch modes are close to the Brillouin zone center, and provided that the mechanical frequency is within the experimental range, the two modes (i.e.~polarizations) are not coupled and one can resort to a single-optical-mode perturbation theory. We summarize here the main results of our QNMPT, while the complete derivation can be found in Suppl.~Sect.~IV. Indicating with $\vb*{k}$ the photonic Bloch mode wavevector and with $\omega_{\vb*{k}, 0}$ the dispersive optical resonance frequency for the unperturbed metasurface, it can be found that the time-dependent complex resonance frequency for the perturbed metasurface is
\begin{eqnarray}
\omega_{\vb*{k}} (t) = \omega_{\vb*{k}, 0}\, [\,  1 & + & g_{Re}\, \delta_M\, \cos(\Omega t - \Phi_{Re}) \nonumber \\ & + & i\, g_{Im}\, \delta_M\, \cos(\Omega t - \Phi_{Im})\,  ]
\label{eq:main:omegakt}
\end{eqnarray}
where $\Omega$ is the mechanical frequency, $\delta_M$ is the spatio-temporal maximum of the displacement field, $\Phi_{Re(Im)}$ are phase terms defined in Suppl.~Sect.~IV, and the couplings $ g_{Re(Im)}$ are given by the following formulas:
\begin{eqnarray}
g_{Re} & = & f_{\vb*{K}}\ g_{Re}^{(0)} =  \frac{f_{\vb*{K}}}{\omega_{\vb*{k}, 0}} \sqrt{ \Re( \omega_{\vb*{k}, 0} \Delta_{Re} )^2 +  \Re( \omega_{\vb*{k}, 0} \Delta_{Im} )^2} \nonumber \\
g_{Im} & = & f_{\vb*{K}}\ g_{Im}^{(0)} =  \frac{f_{\vb*{K}}}{\omega_{\vb*{k}, 0}} \sqrt{ \Im( \omega_{\vb*{k}, 0} \Delta_{Re} )^2 +  \Im( \omega_{\vb*{k}, 0} \Delta_{Im} )^2} \nonumber
\end{eqnarray}
Here, $f_{\vb*{K}}$ is a real, adimensional quantity that describes the effectiveness of the coupling between the mechanical mode of Bloch wavevector $\vb*{K}$ and the electromagnetic field. In principle, for an infinitely extended metasurface,  $f_{\vb*{K}} = \delta(\vb*{K})$, i.e.~$f_{\vb*{K}}$ is a Dirac delta centered at the origin (the $\Gamma$ point) of the Brillouin zone, meaning that only those phonons can induce optomechanical modulation. However, we have postulated an ``effective coupling region'' scheme (see Suppl.~Sect.~IVa) such that, in real space, QNMPT integrals can be limited to a finite portion of the metasurface area. This assumption enables coupling with mechanical modes at $\vb*{K} \neq 0$, and will be later instrumental to interpret the experimental data. The quantities $\Delta_{Re}$ and $\Delta_{Re}$, both being in general complex, directly originate from the perturbative integrals calculated over the single unit cell for the moving boundary (MB) and photoelastic (PE) effects: $\Delta_{Re} = \Delta_{\mathrm{MB}, Re} + \Delta_{\mathrm{PE}, Re}$ and $\Delta_{Im} = \Delta_{\mathrm{MB}, Im} + \Delta_{\mathrm{PE}, Im}$. We report here the expression for the $\Delta_{\mathrm{MB}, Re}$ term:
\begin{widetext}
\begin{equation}
\Delta_{\mathrm{MB},Re} = - \frac{1}{2D_\mathrm{uc}\, \delta_{\mathrm{ref}}} \int _{\partial \mathcal{D}_\mathrm{uc}}\mathrm{d}S\ \Re[\vb*{n} \cdot \vb*{u}^P_{\vb*{K}, N}]\  \bigg(  \Delta \eps \vb*{E}_{\parallel, \vb*{-k}}^P\cdot \vb*{E}_{\parallel, \vb*{k}}^P  - \Delta \eta \vb*{D}_{\perp, \vb*{-k}}^P \cdot \vb*{D}_{\perp, \vb*{k}}^P \bigg).
\label{eq:main:deltaMBRe}
\end{equation}
\end{widetext}
The integral is performed over the boundary surface that delimits the dielectric material unit cell (region $\partial \mathcal{D}_\mathrm{uc}$) and the surrounding air. $\vb*{n}$ is the outgoing unit vector normal to the boundary; $\Delta \eps = \eps_{\mathrm{GaAs}} - 1$ and $\Delta \eta = 1/\eps_{\mathrm{GaAs}} - 1$. $\vb*{u}^P_{\vb*{K}, N}$ is the periodic part of the mechanical Bloch wavefunction of Bloch wavevector $\vb*{K}$ and mode index $N$. $\vb*{E}_{\parallel, \vb*{k}}^P$ is the periodic part of the $\vb*{k}$-Bloch-wavevector electric field eigenfunction, projected parallelly to the boundary surface. Analogously, $\vb*{D}_{\perp, \vb*{k}}^P$ is the displacement field projected perpendicularly to the boundary surface. Notice that the $\vb*{E}$ and $\vb*{D}$ fields come in pairs with $\vb*{k}$ and $-\vb*{k}$, without complex conjugation: this is a key feature of QNMPT, where the ordinary orthogonality relations of Bloch functions no longer holds. Rather, one has to rely on right and left eigenvectors of the Maxwell equation operator, that, in the case of periodic systems, consist of the $(\vb*{k},-\vb*{k})$ pair. Finally, $D_\mathrm{uc}$ is a normalization integral that in general assumes complex values, and $\delta_{\mathrm{ref}}$ is a reference displacement (originating from the mechanical field numerical solver). The term $\Delta_{\mathrm{MB}, Im}$ is given by Eq.~(\ref{eq:main:deltaMBRe}) with $\Re[\vb*{n} \cdot \vb*{u}^P_{\vb*{K}, N}]$ replaced by $\Im[\vb*{n} \cdot \vb*{u}^P_{\vb*{K}, N}]$. The real and imaginary PE terms have a similar formulation, whose explicit expressions have been reported in the Suppl.~Sect.~IVa.

The presence of complex quantities in Eq.~(1) implies that both the real and imaginary part of $\omega_{\vb*{k}}$ are modulated in time, i.e.~that both resonance frequency and quality factor of the dispersive resonance can be tuned optomechanically. Noticeably, since in principle $\Phi_{Re} \neq \Phi_{Im}$ , the trajectory of $\omega_{\vb*{k}} (t)$ in the complex plane is an elliptical curve. Moreover, in the terminology of quantum optomechanics, our system exhibits in principle both dispersive and dissipative coupling. It will however turn out from the numerical calculations that $g_{Im}/g_{Re} \approx 10^{-2}$, thus the dissipative contribution is negligible, at least within the approximations mentioned so far. 

We will now derive how the modulation of the resonance frequency can be detected by means of a spectroscopic modulation experiment. The key theoretical assumption is that the metasurface transmittance follows a Fano lineshape, as predicted by single-mode coupled-mode theory \cite{Zanotto2018}:
\begin{equation}
T(\omega) = \left| \tau + \frac{\Im(\omega_{\vb*{k}})\, (\rho + i \tau)}{\omega - \Re(\omega_{\vb*{k}}) - i \Im(\omega_{\vb*{k}})} \right|^2 
\label{eq:main:fano}
\end{equation}
Here, $\rho$ and $\tau$ are the off-resonance reflection and transmission coefficients $\rho^2+\tau^2=1$, which are reasonably unaffected by the mechanical motion. The temporal modulation of $T$ stems from the dependence of $\omega_{\vb*{k}}$ upon the mechanical action. Employing Eq.~(\ref{eq:main:omegakt}), the approximation $g_{Im} \ll g_{Re}$, and making the wavelength explicit, the following expression is obtained:
\begin{equation}
T(\lambda,t) = T_0 (\lambda) + \frac{1}{2} \Delta T (\lambda) \cos \left( \Omega t - \Phi_{Re} + \pi \right)
\label{eq:main:Tmod}
\end{equation}
where the static term $T_0 (\lambda)$ is obtained from Eq.~(\ref{eq:main:fano}) by replacing $\omega_{\vb*{k}}$ with the static frequency $\omega_{\vb*{k},0}$. The modulation term amplitude is given by
\begin{equation}
\Delta T (\lambda) =  2\, \frac{\ud T_0}{\ud \lambda} \bigg| _{\lambda_{\vb*{k} ,0}} \lambda_{\vb*{k}, 0}\, g_{Re}\, \delta_M
\label{eq:main:DeltaT}
\end{equation} 
where $\lambda_{\vb*{k}, 0} = 2 \pi c /\omega_{\vb*{k},0}$. Equations (\ref{eq:main:Tmod}) and (\ref{eq:main:DeltaT}) are the main link between theory and experiment: the RF component of the transmitted light at $\Omega$ is in fact proportional to $|\Delta T|$. In our experiment, where we analyzed the transmitted light using a high frequency detector coupled with a spectrum analyzer, we were unable to measure the phase $\Phi_{Re}$; this can however be accessed by using a measurement scheme employing a vector network analyzer.

In the first part of our experimental study, we analyzed the angle-dependent modulation effect at fixed modulation frequency. In detail, we employed a metasurface-IDT pair where the IDT (and, consequently, the RF source) was matched to $\Omega/2\pi = 1102.7\ \MHz$. The beam of a tunable laser with wavelength spanning over the telecom C-band was shone on the metasurface, and the static transmittance $T_0 (\lambda)$ (i.e., that with RF off) was collected. Switching on the RF signal with a power of -21 dBm and sweeping the laser wavelength, we collected the modulation spectrum $|\Delta T (\lambda)|$. Figure 2a illustrates the static and dynamic spectral behavior of the nonlocal resonance under analysis. 
\begin{figure*}
\centering
\includegraphics[width = \textwidth]{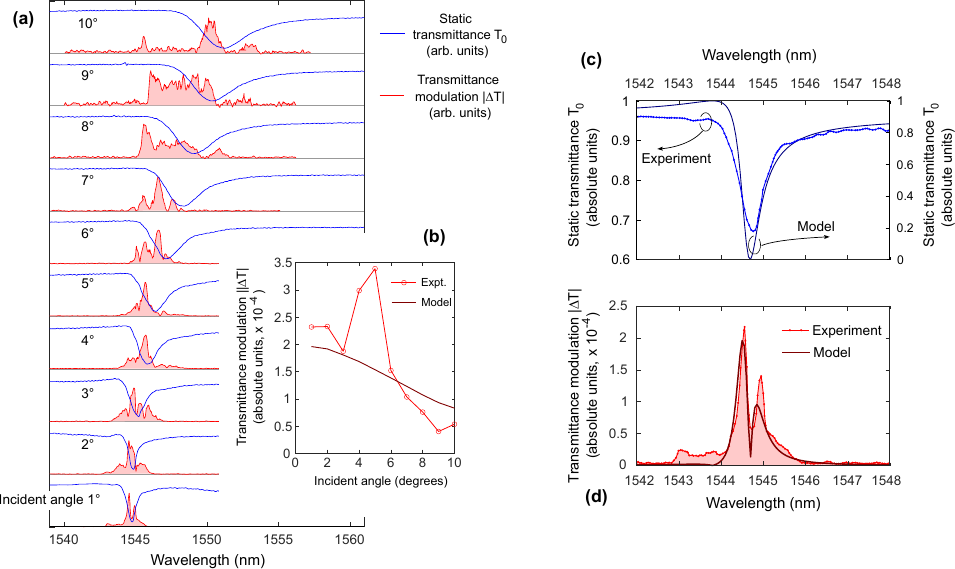}
\caption{Angular and spectral properties of the metasurface modulation. (a), angle-dependent transmittance modulation spectra, compared with static transmittance spectra, indicate that modulation occurs in the vicinity of the angularly dispersive resonance. (b) angular trend of the transmittance modulation spectral maximum. (c) and (d), zoomed view of static and modulated transmittance spectra at $1^{\circ}$ incidence angle.}
\end{figure*}
Here, $T_0$ very well matches the Fano lineshape function, with central wavelength and linewidth drifting with the incident angle. In agreement with theoretical modeling (see Suppl.~Sect.~III), the resonance Q-factor varies with angle from slightly below 3000 to approximately 450. The dynamic response, i.e.~the transmittance modulation amplitude $|\Delta T (\lambda)|$, is reported as red curves, offset in order to be compared with the respective $ T_0 (\lambda)$. It can be noticed that the  $|\Delta T|$ vanishes except for the wavelength range that surrounds the Fano dip of $ T_0 $; the width of such range increases with $\theta$ according to the decrease of the static transmittance Q-factor. While the spectra in Fig.~2a are each normalized to its maximum in order to improve the clarity of the representation, in Fig.~2b we report modulation data expressed in absolute units. Here, we plot, as a function of $\theta$, the maximum of each spectrum $|\Delta T (\lambda)|$. Theory and model match in the overall trend of decreasing $|\Delta T|$ for increasing $\theta$, except for the peak observed at $\theta= 4^{\circ}$ and  $5^{\circ}$. We however attribute this observation to the error induced by the re-alignment procedure ($x-y$ sample position) after an angular rotation, in conjunction with a certain degree of spatial inhomogeneity (see later). Besides this mismatch, we found an excellent agreement in the order of magnitude of the modulation amplitude. In our coupling model we indeed have only a single fitting parameter, that is, the mechanical displacement $\delta_M$; the curves in Fig.~2b and 2d have been obtained using $\delta_M = 0.37\ \picom$, to be compared with $\delta_M \approx 0.28\ \picom$ deduced independently via a procedure detailed in Suppl.~Sect.~X.

In a further analysis, we have focused on the spectral lineshape of the static and dynamic transmittance observed for $\theta = 1^{\circ}$. In Figs.~2c and 2d we show, respectively, the spectra of $T_0$ and $|\Delta T|$, in absolute units, following from experiment and model. Both experimental spectra match very well the theoretical counterparts except for minor discrepancies. In the spectra of $T_0$, the experimental peak has indeed a minimum value of $\approx 0.67$ as opposed to the value of 0 observed in the theory; this effect is attributed to scattering losses induced by imperfections in the metasurface periodicity and hole shape. Moreover, in our model, the spectral feature of $|\Delta T (\lambda)|$ is a double-peaked curve originating from the wavelength derivative of $T_0(\lambda)$. While this aspect is evident in the experimental data for $\theta = 1^{\circ}$, more complex features are observed in the spectra of $|\Delta T (\lambda)|$ for larger values of the incident angle. We attribute this phenomenon to two effects. First, the effective increment of the spot size occurring for larger incident angles leads to effectively probing of a larger area of the metasurface. Indeed, as follows from the analysis of a spatial mapping of the modulation (see Suppl.~Sect.~IX), the metasurface response is not homogeneous, featuring instead fabrication-induced disorder that may be contributing to the aforementioned effect. Second, the contribution of higher order scattering mechanisms could be relevant. Indeed, it is known that even a small amount of roughness enables the population of optical modes at the same laser frequency but at $\vb*{k}$’s different from that of the directly excited mode \cite{Regan2016}. At the operating frequencies several optical modes exist (see Suppl.~Sect.~III), and in particular at larger $\theta$ we found a region rich of other modes, which could lead to a more efficient inter-modal scattering mechanism. The modulated, re-scattered optical energy could be eventually re-coupled to the free space and experimentally detected via the main optical mode, also without perfect resonance matching. 
 
In the second part of our experimental study, we focused on the frequency response of the modulation, in order to gain insight over the bandwidth capabilities of our system, and to obtain a broader picture about the optomechanical coupling mechanisms behind the reported nonlocal metasurface modulation. To this aim we fixed the incident angle to $\theta = 1^{\circ}$ (smaller angles would have led to light backscattering into the laser with consequent instabilities) and measured the $|\Delta T|$ on many metasurface-IDT pairs. We report in Fig.~3a the RF scattering parameter $S_{11}$, whose dips indicate the excitation of a SAW emitted from the IDTs.
\begin{figure*}
\centering
\includegraphics[width = \textwidth]{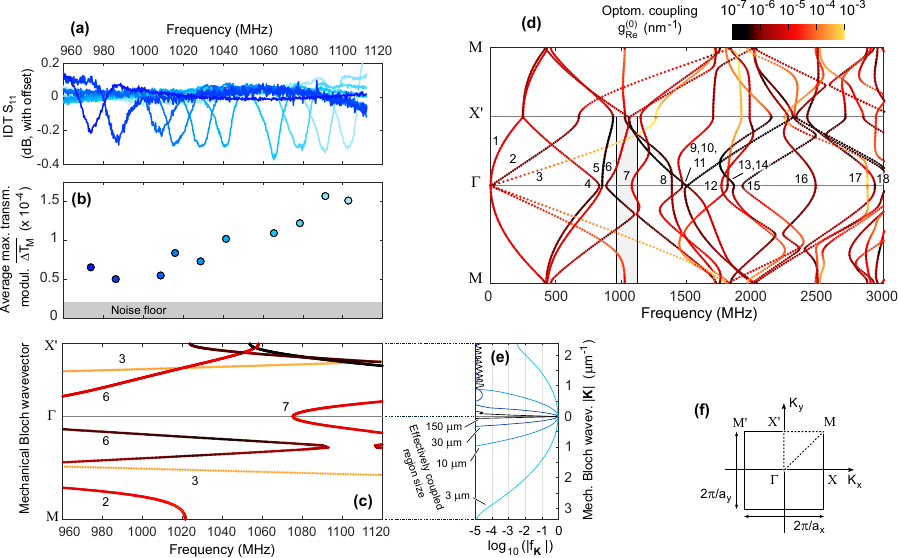}
\caption{Frequency response of modulation and relation with mechanical band structure. (a), RF reflection spectrum of interdigitated transducers (IDTs) with different center frequencies integrated in the chip. (b), average maximum transmittance modulation (see text for the definition) collected from the metasurfaces coupled to the IDTs. The metasurfaces are nominally identical, only IDTs have different pitches. (c-d), calculated mechanical band structure and optomechanical coupling $g_{Re}^{(0)}$ (see Eq.~2). Modes at $\Gamma$ are labeled with increasing numbers. This numbering scheme is used to identify, when no ambiguity occurs, also modes far from $\Gamma$. Panel (c) is a zoomed version of the grey region in (d). (e), coupling factor $f_{\vb*{K}}$ (see Eq.~2) as a function of effectively coupled region size and of mechanical Bloch wavevector modulus. (f), representation of the Brillouin zone for the slightly rectangular crystal constituting the metasurface.}
\end{figure*}
After such a characterization of IDTs, we excited each IDT at the appropriate $\Omega$, collecting $\overline{\Delta T_M} \equiv \left\langle \max_{\lambda} |\Delta T (\lambda)| \right\rangle$, where $\left\langle \cdot \right\rangle$ indicates the spatial averaging over the area of the metasurface facing the active IDT. The measured $\overline{\Delta T_M}$ are plotted in Fig.~3b. Here, it can be noticed that an optomechanical response is observed for all the driving frequencies, with a clear monotonically increasing trend. The presence of a finite $\overline{\Delta T_M}$ for each of the investigated $\Omega$’s means that there is a nonzero coupling between the optical mode and some mechanical mode at the specific $\Omega$. The exact determination of which mechanical mode is involved is not straightforward, for two reasons. First, as recently reported \cite{Zanotto2022}, the mechanical wave undergoes strong scattering processes while crossing the interface between the exterior and the interior of the metamaterial. As a result, the whole set of Bloch modes at $\Omega$ are populated (isofrequency set, see Suppl.~Sect.~VII), leading to a mechanical displacement that does not originate from a pure Bloch modal function $\vb*{u}_{\vb*{K}}$. Second, fabrication-induced disorder can lead to a deviation from the Bloch mode picture, as disorder is known to promote the formation of states in the bandgaps (Urbach tails) \cite{John1987}. Since an ab-initio treatment of those two effects is presently unfeasible, we will interpret the data in Fig.~3b semi-quantitatively, relying on the mechanical band structure of the perfectly periodic metasurface illustrated in Figs.~3c and 3d. The former is a zoomed version of the latter, as indicated by the gray rectangle in Fig.~3d. In both Figs.~3c-d, the Bloch modes are colored according to their optomechanical coupling $g_{Re}^{(0)}$, see the color scale above Fig.~3d. Bloch modes are also labeled by numbers, whose sequence follows from the order of modes at the $\Gamma$ point. We will use the same numbering also to label modes far from the zone center, for those branches that can be followed and unambiguously identified. Notice that in the spectral region of the fabricated IDTs (960-1120 MHz) only mode 7 extends to the $\Gamma$ point, with a minimum at $\Omega/2 \pi = 1075\ \MHz$; its wavefunction at the $\Gamma$ point was depicted in Fig.~1d. Outside from $\Gamma$, other modes exist in the 960-1120 MHz region, with $g_{Re}^{(0)}$ spanning several orders of magnitude. Modes 2 and 7, as well as the $\Gamma$-X’ branch of mode 6, have a mid-range $g_{Re}^{(0)} \approx 5 \cdot 10^{-5}$, while mode 3 features a large coupling value of $g_{Re}^{(0)} \approx 6 \cdot 10^{-4}$.

Let us first analyze the origin of a finite $\overline{\Delta T_M}$ for $\Omega/2 \pi < 1075\ \MHz$, recalling that $\overline{\Delta T_M} \propto \Delta T \propto g_{Re} = f_{\vb*{K}} \cdot g_{Re}^{(0)}$. Two main mechanisms can be responsible for this phenomenon: first, it can originate from tail states of mode 7, i.e.~eigenmodes occurring below the band minimum due to disorder effects. The existence of finite displacement close to the $\Gamma$ point for $\Omega/2 \pi < 1075\ \MHz$ is confirmed by displacement interferometry, as discussed in Suppl.~Sect.~VII. Moreover, a contribution from mode 3 - which we know to be populated, see again Suppl.~Sect.~VII - could also occur. However, this requires that $f_{\vb*{K}}$ has a non-negligible value for $|\vb*{K}| \approx 1.5 \div 1.8\ \um^{-1}$, that are the typical momentum values of mode 3 in the selected frequency range (see Fig.~3c). Within the most immediate hypothesis that the effectively coupled region had a size equal to the laser beam spot, i.e.~$30\ \um$, $f_{\vb*{K}}$ would be such that $g_{Re}$ is suppressed with respect to $g_{Re}^{(0)}$ by at least five orders of magnitude (Fig.~3e). The situation would be even worse if the total metasurface size ($150\ \um$) was employed. Nonetheless, a contribution to the modulation from mode 3 can appreciably occur if the effectively coupled region has a size of the order of $3\ \um$, which means $f_{\vb*{K}} \approx (0.5 \div 1)\cdot 10^{-1}$ in the $|\vb*{K}| \approx 1.5 \div 1.8\ \um^{-1}$ range, and hence, considering the $g_{Re}^{(0)}$ values and Eq.~(\ref{eq:main:DeltaT}), a modulation level consistent with the experiment. We believe that such a small effectively coupled region size is reasonable, as it may originate from the effect of spatial inhomogeneities, that limits optomechanical coherency length and prevents it from extending out of micro-metric sized ``islands'' (see also Suppl.~Sect.~IX). Interestingly, the finite coupling region approach justifies also the observed $\Delta T$ above 1075 MHz. Indeed, the $\approx 3\ \um$ coupling region implies that $f_{\vb*{K}}$ is a smooth function, and of order unity, in the $|\vb*{K}| < 0.5\ \um^{-1}$ interval. Since this interval is relevant for mode 7 (see Fig.~3c), the mentioned behavior of $f_{\vb*{K}}$ eventually allows for significant optomechanical coupling with all frequencies in the 1075-1105 MHz range. In particular, the increasing trend of  $\overline{\Delta T_M}$ suggests that coupling with mode 7 is present in addition with coupling with mode 3. Indeed, spanning from 960 and 1120 MHz, mode 3 has a rather constant value of $g_{Re}^{(0)}$, while its $|\vb*{K}|$ is increasing: if the optomechanical effect originated only from mode 3, its strength would have decreased with increasing frequency due to the $f_{\vb*{K}}$ trend. 

The final part of our study is a numerical analysis on the role of mirror plane symmetries on the optomechanical effect. As well known in solid-state physics, symmetries imply selection rules that eventually dictate allowed and forbidden processes involving multiple states. In our case, symmetries pose strong constraints on the finiteness of the coupling integrals $\Delta_{Re(Im)}$  and eventually of the coupling constants $g_{Re(Im)}$ and experimentally measurable modulation $\Delta T$. The first observation is that, for zero slant angle $\beta$, our unit cell is invariant for the mirror symmetry operations $\hat{\sigma}_x$, $\hat{\sigma}_y$ and $\hat{\sigma}_z$, while for a finite $\beta$ only $\hat{\sigma}_x$ and $\hat{\sigma}_y$ remain symmetries of the system. Since, however, in our sample $\beta$ is small, we can start studying the structure with $\beta = 0$ and classify the mechanical eigenmodes and their couplings $g_{Re(Im)}^{(0)}$ according to the mode eigenvalues with respect to the operators $\hat{\sigma}_{x,y,z}$. Following the analysis outlined in Suppl.~Sect.~V, it turns out that, for $\vb*{K} = (0,0)$ and $\vb*{k} = (0,0)$, the integral in Eq.~(3) can be written as
\begin{widetext}
\begin{equation}
\Delta_{\mathrm{MB},Re} = - \frac{(1+\sigma_x)(1+\sigma_y)(1+\sigma_z)}{2D_{1/8}\, \delta_{\mathrm{ref}}} \int _{\partial \mathcal{D}_{1/8}}\mathrm{d}S\ \Re(\vb*{n} \cdot \vb*{u})\  \bigg(  \Delta \eps\, \vb*{E}_{\parallel}\cdot \vb*{E}_{\parallel} - \Delta \eta\, \vb*{D}_{\perp} \cdot \vb*{D}_{\perp} \bigg)	
\label{eq:main:DeltaMBsymm}
\end{equation}
\end{widetext}
where $\sigma_{x,y,z}$ are the eigenvalues of the operators $\hat{\sigma}_{x,y,z}$ for the mechanical wavefunction $\vb*{u}$, and the integral is taken over one appropriately chosen eighth of the unit cell. Analogous formulas hold for $\Delta_{\mathrm{MB},Im}$ and $\Delta_{\mathrm{PE},Re(Im)}$. An immediate consequence of Eq.~(\ref{eq:main:DeltaMBsymm}) is that it is sufficient that only one among $\sigma_{x,y,z}$ equals -1 in order to have $\Delta_{Re} = \Delta_{Im} = 0$ (recall that the eigenvalues of mirror symmetry operators can only take $\pm 1$ values). Interestingly, the eigenvalue of the mirror symmetry operation for the e.m.~wavefunction does not play any role since the integral is quadratic in the e.m.~fields. The predictive power of Eq.~(\ref{eq:main:DeltaMBsymm})) can be appreciated looking at Fig.~4a, where we plot the values of $g_{Re}^{(0)}$ as blue bars for various mechanical modes at $\vb*{K} = (0,0)$ of the $\beta = 0^{\circ}$ structure. 
\begin{figure*}
\centering
\includegraphics[width = \textwidth]{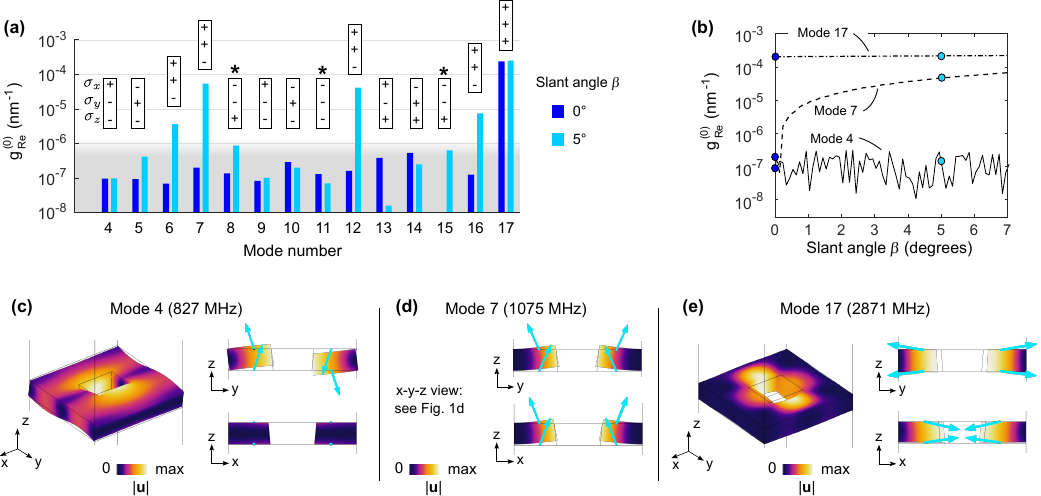}
\caption{Symmetry analysis of optomechanical coupling. (a), optomechanical coupling for a set of mechanical modes (see Fig.~3d for the numbering), calculated for two values of the hole slant angle (see Fig.~1c). In the case $\beta = 0^{\circ}$, finite coupling occurs if all among $\sigma_{x,y,z}$ equal +1. In the case $\beta = 5^{\circ}$, finite coupling occurs if both $\sigma_x$ and $\sigma_y$ equal +1. Bars extending only in the gray region are numerical zeros. The modes marked with an asterisk require in principle two-mode (degenerate) perturbation theory, thus the results may be inaccurate. (b), dependence of the optomechanical coupling upon the symmetry-breaking parameter $\beta$, for modes with different symmetries. Mode 4 never couples because it has $\sigma_y = -1$. Mode 7 couples only for $\beta \neq 0^{\circ}$ because it has both $\sigma_x$ and $\sigma_y$ equal to +1 but $\sigma_z$ equal to -1. Mode 17 always couples, independently of $\beta$, because it has all $\sigma_{x,y,z}$ equal to +1. Panels (c-e) represent the mechanical Bloch mode profile for the three modes analyzed in (b). }
\end{figure*}
Also the photonic $\vb*{k}$ is chosen to be zero. Mechanical modes 1-3 are not considered here since they have zero frequency at the $\Gamma$ point, leading to numerical instabilities. From Fig.~4a it is evident that only mode number 17 has a finite coupling, i.e. a level of $g_{Re}^{(0)}$ that emerges out of the grey region, which represents a numerical zero background (originating from tiny asymmetries in the FEM solver mesh). Notably, mode 17 is the first mode without -1 among the symmetry eigenvalues (refer to Fig.~3d for the numbering of eigenmodes), thus confirming the validity of Eq.~(\ref{eq:main:DeltaMBsymm}). When the condition $\beta = 0^{\circ}$ is relaxed, Eq.~(\ref{eq:main:DeltaMBsymm}) would be replaced by its analogous without the $(1+\sigma_z)$ factor and with the integral performed over one fourth of the unit cell. Thus, the condition to have finite coupling only requires that $\sigma_x$ and $\sigma_y$ both equal +1. In this framework, in addition to mode 17, also modes 6, 7, 12 and 16 should have $g_{Re} \neq 0$: this is exactly what resulted from numerical calculations for $\beta = 5^{\circ}$ (Fig.~4a, cyan bars). We also performed a fine sweep over $\beta$, obtaining the trends for $g_{Re}^{(0)}$ reported in Fig.~4b. Here, one can notice that: (i) mode 4 has a coupling that always remains in the numerical zero region, (ii) mode 7 rapidly acquires a finite value upon increasing $\beta$, (iii) mode 17 has a coupling almost independent of $\beta$. This confirms the crucial role of symmetry eigenvalues of the mechanical modes in the determination of the optomechanical coupling. Moreover, it turns out that a very small deviation from full symmetry, i.e.~a very small value of hole slant angle $\beta$, can be responsible for a significant value of the coupling constant, also for modes that would have zero coupling otherwise. To complete our picture, we plotted in Fig.~4c-e visual representations of the displacement field $\Re(\vb*{u})$ for the three modes analyzed above. Notice that these fields are calculated for the $\beta = 5^{\circ}$ structure, thus in principle they do not have a definite character upon the $\sigma_z$ operation. However, at least visually, a definite character stands out, and such character is exactly that of the $\sigma_z$ eigenvalue determined for the mode of the $\beta = 5^{\circ}$ structure. This supports once more the suitability of symmetry analysis for the study of coupling in slightly perturbed optomechanical slabs. As a final note, we highlight that, according to the analysis detailed in Suppl.~Sect.~Va, the mechanical modes with $\sigma_x = \sigma_y = -1$ (modes 8, 11 and 15) cannot be treated with the single-mode perturbation theory, as they couple the two quasi-degenerate photonic modes falling in our wavelength range. The bars reported in Fig.~4a corresponding to those mechanical modes only take into account the diagonal perturbation elements; thus, the reported $g_{Re}^{(0)}$ values for those modes are subject to corrections. The analysis of this kind of optomechanical degenerate coupling in dispersive metasurfaces is likely a research topic of interest for future studies, possibly leading to useful effects related, for instance, to polarization manipulation.

\section{Discussion and conclusions}
In this work we have shown that, relying on a monolithic piezoelectric dielectric metasurface chip, nonlocal photonic resonances can be electrically modulated at gigahertz frequency. Experiments agree with numerical modeling and with symmetry analysis; similarly to recent reports about chirality \cite{Chen2023}, the tuning of the metasurface holes slant angle appears as a precious tool for metasurface functionality control. In our samples, deviations from perfect periodicity induce spurious scattering, likely responsible for non-ideal modulation lineshapes; however, such nonidealities allow for a more broadband response with respect to what expected for a perfectly periodic metasurface. While our experiments are performed in a purely classical regime, quantum emitters can be easily integrated in the proposed platform, leading to interesting perspectives for quantum optomechanics \cite{Kort-Kamp2021, Balram2016, Balram2022, Stockill2019}. Improving the quality factor of the photonic resonance, for instance relying on bound states in the continuum \cite{Zanotto2022a}, could enable low-power optical modulators, enhanced sensing schemes \cite{Shnaiderman2020} and electrophononics \cite{Akimov2017}. Other applications such as space-time photonics and nonreciprocal devices could benefit from the outcomes of our study \cite{Sohn2018, Kittlaus2021}; in parallel, basic science effects such as quantum Hall effect for light \cite{Fang2019} could possibly find a feasible implementation in our system.

\section*{Acknowledgements}
The authors gratefully acknowledge Chiara Massetti for her contributions in an early stage of this project, as well as for preparing part of Figure 1. Dr. Paulo V. Santos is also warmly acknowedged for the interferometric displacement measurements and for insightful discussions.


\end{document}


\title{Supplementary Material for the article: \\ Gigahertz modulation of a fully dielectric nonlocal metasurface}

\author{Alessandro Pitanti}
\affiliation{CNR-Istituto Nanoscienze, Laboratorio NEST, Piazza San Silvestro 12, 56127 Pisa, Italy}
\affiliation{Paul-Drude-Institut für Festkörperelektronik, Leibniz-Institut im Forschungsverbund Berlin e. V., 5-7, Hausvogteiplatz, Berlin, 10117 (Germany)}

\author{Gaia Da Prato}
\affiliation{Dipartimento di Fisica “E. Fermi”, Università di Pisa, Largo Pontecorvo 3, 56127 Pisa (Italy)}
\altaffiliation{Present address: Kavli Institute of Nanoscience, Department of Quantum Nanoscience, Delft University of Technology, 2628CJ Delft, The Netherlands}

\author{Giorgio Biasiol}
\affiliation{CNR-Istituto Officina dei Materiali, Laboratorio TASC, 34149 Trieste (Italy)}

\author{Alessandro Tredicucci}
\affiliation{Dipartimento di Fisica “E. Fermi”, Università di Pisa and CNR-Istituto Nanoscienze, Largo Pontecorvo 3, 56127 Pisa (Italy)}

\author{Simone Zanotto}
\affiliation{CNR-Istituto Nanoscienze, Laboratorio NEST, Piazza San Silvestro 12, 56127 Pisa, Italy}
\email{simone.zanotto@nano.cnr.it}

\date{\today}

\maketitle

\section{Metasurface fabrication}
The metasurface samples have been fabricated starting from a GaAs wafer, where an epitaxial bilayer consisting of $\alzcq$ (500 nm) and GaAs (220 nm) has been deposited. Next, the hole arrays have been defined by means of e-beam lithography (Zeiss Ultraplus SEM + Raith Multibeam lithography system at 20 keV; AR-P 6200 resist spinned twice at 4000 rpm, thickness 280 nm from profilometer). The pattern has then been transferred to the semiconductor layer by means of ICP-RIE with $\mathrm{Cl}_2$/$\mathrm{BCl}_3$/$\mathrm{Ar}$ gas mixture (Sentech machine). Subsequently, the IDTs have been defined by means of e-beam lithography (Zeiss Ultraplus SEM + Raith Multibeam lithography system at 20 keV; AR-P 669.04 resist spinned at 4000 rpm), thermal evaporation (50 nm Au) and lift-off. In a separate lithography + lift-off process we defined the bus and pads for connections to the IDTs (Durham MagnetoOptics ML3 laser writer; LOR/S1805 resist bilayer; 6/70 Cr/Au thermal evaporation). To obtain the suspended membrane in the metasurface region, the chip has been etched from the back side with a sequence of wet etching steps. First, a fast etching ($\mathrm{H}_2\mathrm{SO}_4$\,:\,$\mathrm{H}_2\mathrm{O}_2$\,:\,$\mathrm{H}_2\mathrm{O}$, 1:8:1) with $\approx 10\ \micron$ per minute etch rate is employed to thin the substrate down to $\approx 50\ \micron$. A resist mask (S1818) is employed in order to protect the chip on the edges, thus allowing for chip handling and subsequent mounting. Second, a slow etching (3:1 solution of citric acid in 30\% hydrogen peroxide; citric acid is prepared in 1:1 weight ratio with water) having $\approx 1\ \micron$ per minute etch rate is employed to remove the remaining GaAs substrate. This solution is selective and stops when the $\alzcq$ layer is reached. In this phase a second resist mask is employed, in order to realize membranes of $\approx 200\ \micron \times 200\ \micron$ size, in correspondence to the metasurface pattern previously realized on the chip front side. Finally, an HF bath (50\% concentration) is employed to remove the $\alzcq$ layer. During the first two wet etching steps the chip is mounted on a glass slide by means of S1818 resist.

\section{Experimental apparatus}
\label{sect:ExpSetup}
The metasurface response, static and dynamic, has been probed on an optical bench consisting of: a tunable laser source, polarizer and waveplate, focusing and collecting lenses (see also Suppl.~Sect.~\ref{sect:SpatialMapping}), slow detector (DC - 1 MHz), fast detector (1 MHz - 2 GHz), oscilloscope, RF generator and RF spectrum analyzer. The $S_{11}$ curves reported in main Fig.~3a were collected with a vector network analyzer. MATLAB-based script were employed to automatize the data collection and to perform data pre-processing. 

\section{Characterization of the nonlocal optical resonance}
\label{sect:characterization}
\begin{figure*}
\centering
\includegraphics[width = \textwidth]{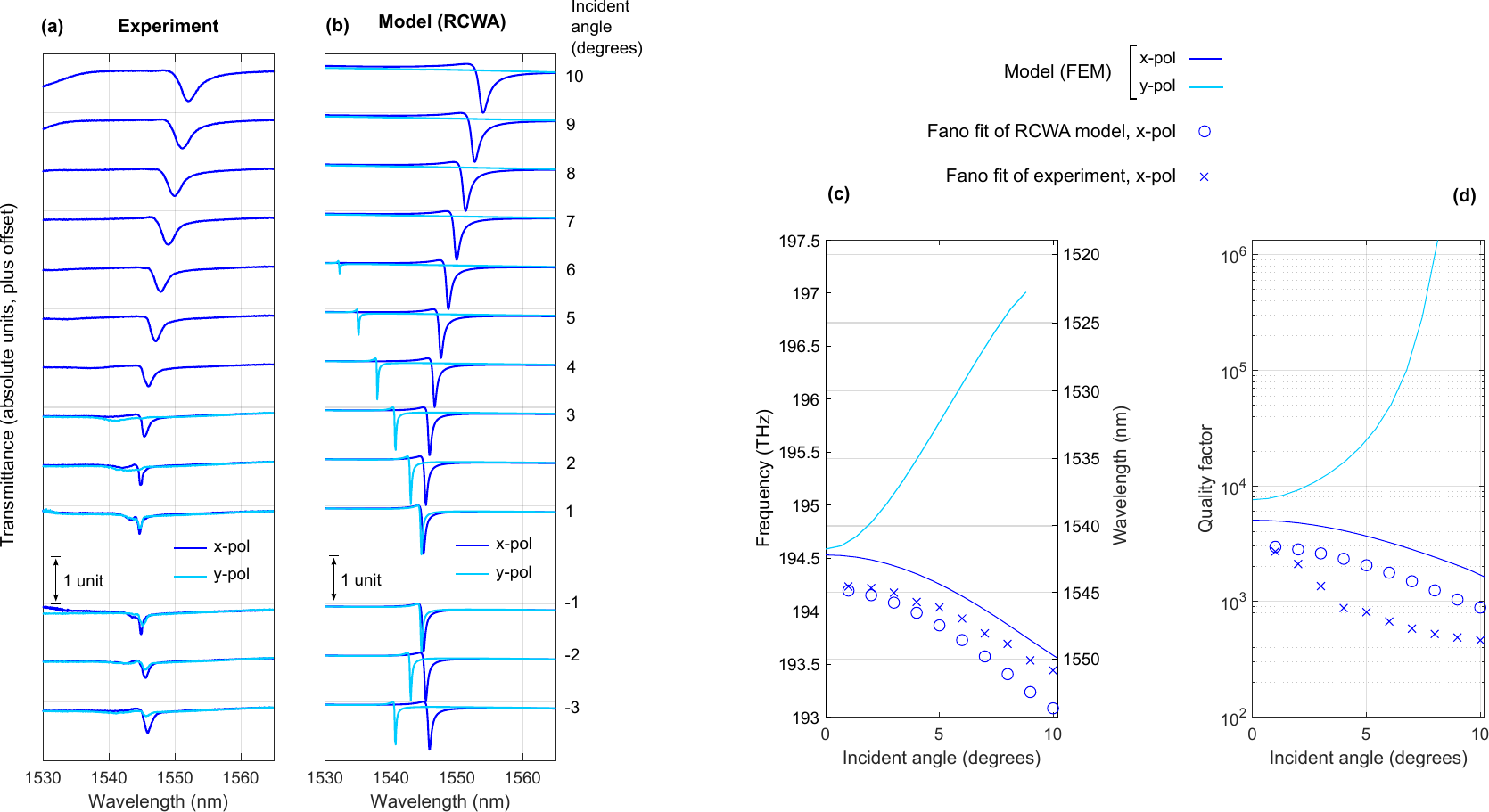}
\caption{Transmission spectra of the metasurface and their analysis.}
\label{fig_staticspectra}
\end{figure*}

The dispersion of the nonlocal optical resonance has been characterized by comparing the results of angle-resolved spectroscopy and those of numerical calculations. Suppl.~Fig.~\ref{fig_staticspectra}a shows the measured transmittance spectra collected for two orthogonal polarization states of the incident light (transmitted light was not polarization-analyzed). The spectrum at $0^{\circ}$ is not reported since it was affected by laser instabilities originating from light backscattered into the laser cavity. Suppl.~Fig.~\ref{fig_staticspectra}b plots the corresponding theoretical spectra, calculated with the rigorous coupled wave analysis (RCWA) method (PPML toolbox for MATLAB\footnote{\href{https://it.mathworks.com/matlabcentral/fileexchange/55401-ppml-periodically-patterned-multi-layer}{https://it.mathworks.com/matlabcentral/fileexchange/55401-ppml-periodically-patterned-multi-layer}\\ \href{https://github.com/zan8simone/PPML---Periodically-Patterned-Multi-Layer}{https://github.com/zan8simone/PPML---Periodically-Patterned-Multi-Layer}}). 
It can be noticed that the theoretically predicted x-polarized resonance is clearly observed in the experiment, while the y-polarized resonance (that has a very narrow theoretical linewidth) is almost invisible in the measurement. The presence of a small secondary peak in the measured x-polarized spectra at $\pm 1^{\circ}$ probably originates from polarization mixing in the sample. Following general knowledge about photonic crystal slab resonances, the peaks have been fitted with Fano lineshapes:
\[
T = \left| \tau + \frac{\gamma_1 (\rho + i \tau)}{\omega - \omega_0 - i (\gamma_1 + \gamma_2)} \right|^2 ,
\]
where $\tau\ \in [0,1]$ is the off-resonance transmission amplitude, $\rho = \sqrt{1-\tau^2}$, $\omega$ ($\omega_0$) are the radiation (resonance) angular frequencies, $\gamma_1$ is the loss rate in the main radiative channel, and $\gamma_2$ is the loss rate in other channels, such as disorder-induced scattering channels. In the formula above, the presence of $\gamma_2$ enables non-fully-contrasted lineshapes, thus allowing to adequately fit experimental spectra. Lack of full contrast can originate from either scattering losses or from the finite hole slant angle.  Concerning this aspect, RCWA, while being a very fast method, has the disadvantage that it cannot account for slanted holes, as opposed for instance to the finite element method (FEM). We hence performed eigenmode analysis with COMSOL FEM: with this method, the results plotted in Suppl.~Fig.~\ref{fig_staticspectra}c-d as continuous lines have been obtained. A good overall agreement with RCWA and experimental data is observed; the slight shift can be indeed attributed to the finite hole slant angle effect on photonic resonances.

\begin{figure*}
\centering
\includegraphics{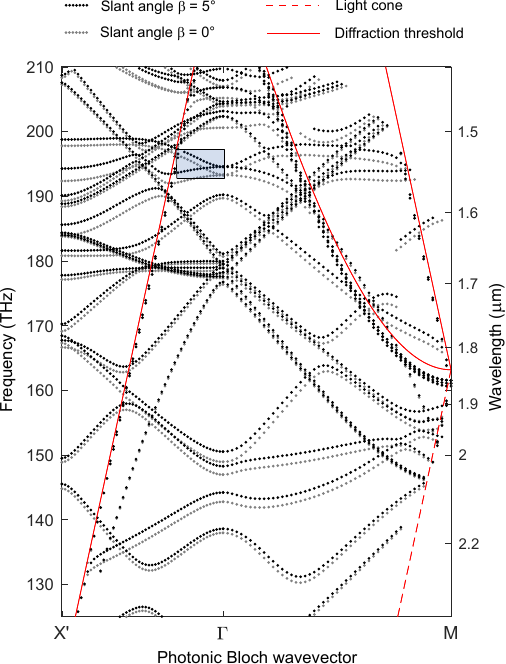}
\caption{Photonic mode structure of the metasurface.}
\label{fig_photonicbands}
\end{figure*}

To get a broader picture on the resonance structure of our metasurface, we calculated with the FEM the photonic Bloch mode dispersion along some of the high symmetry paths in the first Brillouin zone. The result is reported in Suppl.~Fig.~\ref{fig_photonicbands}, where data for slanted and non-slanted holes are reported. 
Consistently with what observed above, the slanted structure has blue-shifted modes with respect to the non-slanted one; moreover, it is interesting to notice that the overall mode structure is conserved, irrespectively of the slant angle. The greyed out region is that corresponding to Suppl.~Fig.~\ref{fig_staticspectra}. It can be noticed that the left side of that region (i.e., that with larger $|\vet k|$'s) is close to a multitude of modes: as discussed in the main text, this could be one reason behind the imperfect modulation lineshapes observed at large incident angles. In Suppl.~Fig.~\ref{fig_photonicbands} only the modes with a Q-factor $>$ 100 have been plotted, to avoid showing spurious solutions. This cutoff may have however introduced some artifacts such as some jumps observable close to the M point. Modes can also be missing in the top-right region of the plot owing to the finite number of solutions calculated by COMSOL.
\section{Quasi-normal-mode perturbation theory}
\label{sect:QNMPT}
We developed a quasi-normal-mode perturbation theory (QNMPT) adapting the results of \cite{Primo2020} to our system. Let's assume that there is only one photonic mode involved; we will justify this later. Starting point is the formula
\begin{equation}
\label{eq:QNM_PT}
\Delta \omega = -\frac{\omega_0}{2} 
\frac{\int\vb*{E}^L\, \Delta\varepsilon\, \vb*{E}^R\, \mathrm{d}V}
{\int\vb*{E}^L\, \varepsilon\, \vb*{E}^R\, \mathrm{d}V}
\end{equation}
where $\omega_0$ is the unperturbed eigenfrequency, $\vb*{E}^L$ and $\vb*{E}^R$ are the left and right electric field eigenfunctions of Maxwell equations, $\varepsilon$ is the unperturbed permittivity, $\Delta\varepsilon$ is the permittivity perturbation, and the integral extends over the whole computational domain, including the perfectly matched layers (PML). 
For a periodic system, it has been shown that left and right eigenfunctions are Bloch modes at opposite Bloch wavevectors, i.e., $\vb*{E}^L = \vb*{E}_{\vet{-k}} = \vb*{E}^P_{\vet{-k}}\ e^{-i\vet k \cdot \vet r_{\parallel}}$ and $\vb*{E}^R = \vb*{E}_{\vet{k}} = \vb*{E}^P_{\vet{k}}\ e^{i\vet k \cdot \vet r_{\parallel}}$, both corresponding to the eigenfrequency $\omega_{\vet k, 0}$. In the expressions above, ``$P$'' labels the periodic part of Bloch wavefunction and $\vet r_{\parallel} = (x,y)$ (we are focusing on systems periodic in two dimensions). With these positions, Suppl.~Eq.~(\ref{eq:QNM_PT}) becomes
\begin{equation}
\label{eq:QNM_PT_periodic}
\Delta \omega = -\frac{\omega_{\vet k, 0}}{2} 
\frac{\int_{\mathcal{D}'_{\mathrm{full}}} \vb*{E}^P_{\vet{-k}}\, \Delta\varepsilon\, \vb*{E}^P_{\vet{k}}\, \mathrm{d}V}
{\int_{\mathcal{D}'_{\mathrm{full}}} \vb*{E}^P_{\vet{-k}}\, \varepsilon\, \vb*{E}^P_{\vet{k}}\, \mathrm{d}V}
\end{equation}
where $\mathcal{D}'_{\mathrm{full}}$ is the full domain, that for an ideal periodic metasurface would be infinitely extended. While the integral at denominator (named $D_{\mathrm{full}}$ in the following) can be reduced to a unit cell, that at numerator operates on a function that is not guaranteed to be periodic, as it depends on the actual form of $\Delta\varepsilon$. Since $\Delta\varepsilon$ inherits the periodicity from the mechanical Bloch mode, the numerator has a period that is the l.c.m.\ of the periods of photonic and mechanic wavefunctions; if those periods are incommensurate, the integration domain is unbounded. We will tackle this problem later on. 
\begin{figure*}
\centering
\includegraphics{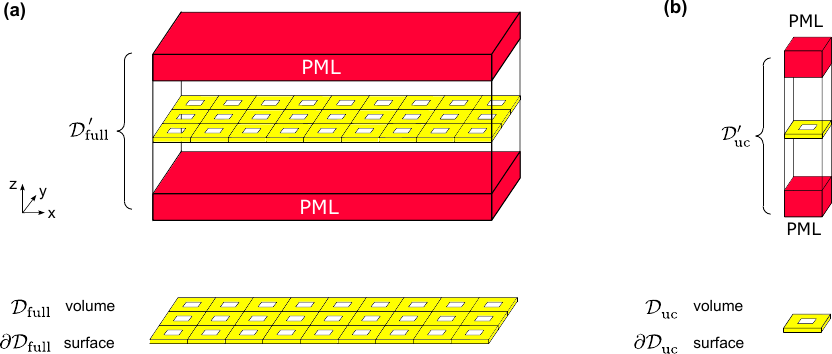}
\caption{Nomenclature for the integration domains. (a), full domains, that would ideally extend to infinity in the $x-y$ plane. (b), unit cell domains.}
\label{fig:domains}
\end{figure*}

Here we will rather focus on giving a more explicit form to Suppl.~Eq.~(\ref{eq:QNM_PT_periodic}), considering that in an optomechanical system the permittivity perturbation originates from two effects: moving boundary and photoelastic. Moving boundary effect calls for a modification of the form of the perturbative integrand $\vb*{E}^L\ \Delta\varepsilon\ \vb*{E}^R$, since the field is discontinuous at the interface between materials. For this purpose, the following form should be rather used:    
\begin{equation}\label{eq:MB_QNM}
\begin{aligned}
\Delta \omega_{\mathrm{MB}} = &
-\frac{\omega_{\vet k, 0}\, \delta_M}{2D_{\mathrm{full}}\, \delta_{\mathrm{ref}}}
\int_{\partial \mathcal{D}_{\mathrm{full}}} \mathrm{d}S\ \vb*{n} \cdot \vb*{\tilde{u}}(\vb*{r}_{\parallel},z,t) \left[   \vb*{E}_{\parallel, \vb*{-k}}^P(\vb*{r}_{\parallel},z)\cdot(\varepsilon_1-\varepsilon_2)  \vb*{E}_{\parallel, \vb*{k}}^P(\vb*{r}_{\parallel},z)\right]\\
&+\frac{\omega_{\vet k, 0}\, \delta_M}{2D_{\mathrm{full}}\, \delta_{\mathrm{ref}}}
\int_{\partial \mathcal{D}_{\mathrm{full}}} \mathrm{d}S\ \vb*{n} \cdot \vb*{\tilde{u}}(\vb*{r}_{\parallel},z,t) \left[  \vb*{D}_{\perp, \vb*{-k}}^P(\vb*{r}_{\parallel},z) \cdot (\varepsilon_1^{-1}-\varepsilon_2^{-1}) \vb*{D}_{\perp, \vb*{k}}^P(\vb*{r}_{\parallel},z)\right].
\end{aligned}
\end{equation}
In substance, the volume integral has been converted into a surface integral, where $\vet n$ is the outward normal vector to $\mathcal{D}_{\mathrm{full}}$ (see Suppl.~Fig.~\ref{fig:domains} for the definition of the domains). $\vb*{\tilde{u}}$ is the mechanical displacement field at a certain time $t$, that we assume to originate from a single Bloch mode:
\[
\vb*{\tilde{u}}(\vb*{r}_{\parallel},z,t) = \Re[\vb*{u}_{\vb*{K},N}(\vb*{r}_{\parallel},z)\, e^{-i\Omega_{\vb*{K},N}t}]= \Re[\vb*{u}_{\vb*{K},N}^P(\vb*{r}_{\parallel},z)\, e^{i\vb*{K}\cdot \vb*{r}_{\parallel}}\, e^{-i\Omega_{\vb*{K},N}t}].
\]
We have introduced here the computational reference factor $\delta_{\mathrm{ref}} = \max_{(\vet r_{\parallel}, z)} \sqrt{ \vb*{u}_{\vb*{K},N}^P \cdot \vb*{u}_{\vb*{K},N}^{P *}  }$, needed to avoid ambiguities on the normalization employed by FEM solvers. $\delta_M$ instead represents the actual maximum displacement of the system under investigation. In Suppl.~Eq.~(\ref{eq:MB_QNM}), $\vet E_{\parallel}$ and $\vet D_{\perp}$ identify, respectively, the projection of $\vet E$ parallel to $\partial \mathcal{D}_{\mathrm{full}}$ and the projection of $\vet D$ perpendicular to $\partial \mathcal{D}_{\mathrm{full}}$; moreover $\varepsilon_1$ and $\varepsilon_2$ are respectively the GaAs and air permittivities. 

Concerning photoelastic effect, one has
\begin{equation}\label{eq:QNM_PE}
\begin{aligned}
\Delta\omega_{\mathrm{PE}} = &\, \frac{\omega_{\vet k, 0}\, \varepsilon_0 \varepsilon_r^2\, \delta_M}{2D_{\mathrm{full}}\, \delta_{\mathrm{ref}}}  \,
\int_{\mathcal{D}_{\mathrm{full}}} \mathrm{d}V \sum_{ijkl} E_{i, \vb*{-k}}^P E_{j, \vb*{k}}^P\,  p_{ijkl}\, \tilde{S}_{kl}
\end{aligned}
\end{equation}
where $\varepsilon_r$ is the GaAs relative permittivity, $p_{ijkl}$ is the photoelastic tensor, and 
\begin{equation}\label{eq:Sdefin}
\begin{aligned}
\tilde{S}_{kl}(\vb*{r}_{\parallel},z,t) &=  \frac{1}{2} \left( \frac{\partial \tilde{u}_{k}}{\partial x_{l}} + \frac{\partial \tilde{u}_{l}}{\partial x_{k}} \right) \\
& = \Re\left[\frac{1}{2}\left( \frac{\partial u_{k, \vb*{K},N}}{\partial x_{l}} + \frac{\partial u_{l, \vb*{K},N}}{\partial x_{k}} \right) e^{-i\Omega_{\vb*{K},N}t}\right]\\
& = \Re\left[\frac{1}{2} \bigg[ \big\lbrace \partial_{l} u_{k, \vb*{K},N}^P + (1- \delta_{l,\,z})\, i K_{l} \, u_{k, \vb*{K},N}^P \big\rbrace + \big\lbrace k \leftrightarrow l \big\rbrace  \bigg] e^{i\vb*{K}\cdot\vb*{r}} e^{-i\Omega_{\vb*{K},N}t} \right]\\
& \equiv \Re\left[S^P_{kl, \vb*{K}, N}\, e^{i\vb*{K}\cdot\vb*{r}} e^{-i\Omega_{\vb*{K},N}t}\right]
\end{aligned}
\end{equation}
is the strain tensor, again defined in terms of the mechanical Bloch mode. In the third line of Suppl.~Eq.~(\ref{eq:Sdefin}), the second term in curly braces is obtained from the first term in curly braces with index exchange. The last line of Suppl.~Eq.~(\ref{eq:Sdefin}) defines the quantity $S^P_{kl, \vb*{K}, N}$, that has the periodicity of the unperturbed metasurface lattice.

\subsection{Effectively coupled region approximation}
When the mechanical Bloch wavevector $\vb*{K}$ is zero, the unit cell of the perturbed structure is the same as in the unperturbed problem. As a consequence, in (\ref{eq:MB_QNM}) and in (\ref{eq:QNM_PE}) one can replace $\mathcal{D}_{\mathrm{full}}$ with $\mathcal{D}_{\mathrm{uc}}$, $\partial\mathcal{D}_{\mathrm{full}}$ with $\partial\mathcal{D}_{\mathrm{uc}}$, and $D_{\mathrm{full}}$ with $D_{\mathrm{uc}} \equiv \int_{\mathcal{D}'_{\mathrm{uc}}} \vb*{E}^P_{\vet{-k}}\, \varepsilon\, \vb*{E}^P_{\vet{k}} \mathrm{d}V$. When instead $\vb*{K} \neq 0$, and the integration domain becomes arbitrarily large for $\vb*{K}$'s belonging to a dense subset of the Brillouin zone, the physical insight suggests that a truncation scheme should be adopted. Various reasons for truncation can be established: the limits of the metasurface itself, i.e.~of the patterned region; the finiteness of the spot size used for probing the optomechanical response (in the case of nonlocal metasurfaces, it not necessarily equal to the region where electromagnetic field is excited); a finite area where fabrication fluctuations are limited below a certain threshold. We will not make here assumption on the reason for truncation (a discussion is given in the main text); rather, we focus on the formulas that can be derived upon limiting the integrals in Eqs.~(\ref{eq:MB_QNM}) and (\ref{eq:QNM_PE}) to $N_x\times N_y$ unit cells. To make the formalism analytically manageable, we assume that $|\vb*{K}|$ is sufficiently smaller than $2\pi/a$ ($a$ being the lattice parameter) such that the function $e^{i\vb*{K}\cdot\vb*{r}}$ can be considered constant in each unit cell (slowly varying envelope approximation). With this assumption, Suppl.~Eq.~(\ref{eq:QNM_PT_periodic}) becomes
\begin{equation}\label{eq:QNM_PT_trunc}
\Delta \omega = \omega_{\vb*{k},0}\, \delta_M\, [f_{cos}(t) \Delta_{Re}-f_{sin}(t) \Delta_{Im} ]
\end{equation}
where $\Delta_{Re}$ is a sum of a MB and a PE contribution: $\Delta_{Re} = \Delta_{\mathrm{MB},Re} + \Delta_{\mathrm{PE},Re}$ and analogously $\Delta_{Im} = \Delta_{\mathrm{MB},Im} + \Delta_{\mathrm{PE},Im}$. Those terms can be calculated by means of single-cell integrals:
\begin{equation}\label{eq:DeltaMBRe}
\Delta_{\mathrm{MB},Re} = - \frac{1}{2D_\mathrm{uc}\, \delta_{\mathrm{ref}}} \int _{\partial \mathcal{D}_\mathrm{uc}}\mathrm{d}S\ \Re[\vb*{n} \cdot \vb*{u}^P]\  \bigg(  \vb*{E}_{\parallel, \vb*{-k}}^P\cdot(\varepsilon_1-\varepsilon_2)  \vb*{E}_{\parallel, \vb*{k}}^P  -  \vb*{D}_{\perp, \vb*{-k}}^P \cdot (\varepsilon_1^{-1}-\varepsilon_2^{-1}) \vb*{D}_{\perp, \vb*{k}}^P \bigg)
\end{equation}
\begin{equation}\label{eq:DeltaPERe}
\begin{aligned}
\Delta_{\mathrm{PE},Re} =  \frac{\varepsilon_0 \varepsilon_r^2}{2D_\mathrm{uc}\, \delta_{\mathrm{ref}}} \int _{\mathcal{D}_\mathrm{uc}} \mathrm{d}V \bigg( &
\ \ E_{x, \vb*{-k}}^P E_{x, \vb*{k}}^P\, \left( p'_{11}\, \Re [ S^P_{x x} ]\, +  p'_{12}\, \Re [ S^P_{y y} ]\, +  p'_{13}\, \Re [ S^P_{z z} ]\, \right) +  \\
&\ \ E_{y, \vb*{-k}}^P E_{y, \vb*{k}}^P\, \left( p'_{12}\, \Re [ S^P_{x x} ]\, +  p'_{11}\, \Re [ S^P_{y y} ]\, +  p'_{13}\, \Re [ S^P_{z z} ]\, \right) +  \\
&\ \ E_{z, \vb*{-k}}^P E_{z, \vb*{k}}^P\, \left( p'_{13}\, \Re [ S^P_{x x} ]\, +  p'_{13}\, \Re [ S^P_{y y} ]\, +  p'_{33}\, \Re [ S^P_{z z} ]\, \right) +  \\
& \left( E_{y, \vb*{-k}}^P E_{z, \vb*{k}}^P\, + E_{z, \vb*{-k}}^P E_{y, \vb*{k}}^P \right)\, 2 p'_{44}\, \Re [ S^P_{y z} ]\ \ + \\
& \left( E_{x, \vb*{-k}}^P E_{z, \vb*{k}}^P\, + E_{z, \vb*{-k}}^P E_{x, \vb*{k}}^P \right)\, 2 p'_{44}\, \Re [ S^P_{x z} ]\ \ + \\
& \left( E_{x, \vb*{-k}}^P E_{y, \vb*{k}}^P\, + E_{y, \vb*{-k}}^P E_{x, \vb*{k}}^P \right)\, 2 p'_{66}\, \Re [ S^P_{xy} ] \ \bigg)  .
\end{aligned}
\end{equation}
The terms $\Delta_{\mathrm{MB/PE},Im}$ are analogous to (\ref{eq:DeltaMBRe}) and (\ref{eq:DeltaPERe}) with all $\Re$'s replaced by $\Im$'s. In Suppl.~Eq.~(\ref{eq:DeltaPERe}), one finds the photoelastic coefficients in Voigt notation for GaAs, oriented such as the $[110]$ and $[1\bar{1} 0]$ crystallographic directions are aligned respectively to the $x$ and $y$ directions of the global coordinate system. Numerically, $p'_{11} = -0.2245$, $p'_{12} = -0.0805$, $p'_{13} = -0.1400$, $p'_{33} = -0.1650$, $p'_{44} = -0.0720$, $p'_{66} = -0.0125$ (ref.~\cite{Yariv1984, Liu2022}). 
The factor $f_{cos}(t)$ is defined as following:
\begin{equation}\label{eq:fcos}
f_{cos}(t) =  \frac{1}{N_x N_y}\sum_{m,l}\cos\left[(a_x\, K_x \,m + a_y\, K_y \, l)-\Omega t\right],
\end{equation}
and $f_{sin}(t)$ analogously replacing the cosine with sine. Here, $K_x=2 \pi/a_x$ and $K_y=2 \pi/a_y$. We have also dropped the mechanical Bloch mode labels $[ {\vb*{K},N} ]$ from $\Omega$ and $\vet u^P$ for notation simplicity. 

It can be however argued that the truncation scheme treated above is too abrupt, and that some smoothing scheme should be employed instead. Based on general physical insight, we introduce in Suppl.~Eq.~(\ref{eq:fcos}) a weight factor drawn from a normalized Gaussian distribution:
\begin{equation}
f_{cos}^{sm}(t) =  \sum_{m,l} \alpha_{m,l}\cos((a_x\, K_x \,m + a_y\, K_y \, l)-\Omega t).
\end{equation}
\begin{equation}\label{eq:alpha_m_l}
\alpha_{m,l} = \frac{\exp\left\{-\left[\left(\frac{a_x m} {\sigma^{\mathrm{eff}}_x}\right)^2+\left(\frac{a_y l} {\sigma^{\mathrm{eff}}_y}\right)^2\right]\right\}}
{\sum_{m',l'}\exp\left\{-\left[\left(\frac{a_x m'} {\sigma^{\mathrm{eff}}_x}\right)^2+\left(\frac{a_y l'} {\sigma^{\mathrm{eff}}_y}\right)^2\right]\right\}},
\end{equation}
with $\sigma^{\mathrm{eff}}_{x,y}$ that define the effectively coupled region size. As a last step we perform further algebraic manipulations to make the temporal dependence of $\Delta \omega$ more explicit. 
By exploiting trigonometric identities and the fact that for any $(m,l)$ pair it exists a $(-m,-l)$ pair, we can write
\begin{equation}\label{eq:QNM_temporal}
\begin{aligned}
\Delta\omega(t) = &\, \omega_{\vb*{k},0}\, \delta_M\, [f_{cos}^{sm}(t) \Delta_{Re}-f_{sin}^{sm}(t) \Delta_{Im} ] \\
= &\,
\omega_{\vb*{k}, 0}\, \delta_M \sum_{m,l}\alpha_{m,l} \left[ \Delta_{Re}\cos(a_x K_x m + a_y K_y l)- \Delta_{Im} \sin(a_x K_x m + a_y K_y l) \right] \cos (\Omega t)\\
&+ \omega_{\vb*{k}, 0}\, \delta_M  \sum_{m,l}\alpha_{m,l} \left[ \Delta_{Re} \sin(a_x K_x m + a_y K_y l)+\Delta_{Im} \cos(a_x K_x m + a_y K_y l) \right] \sin (\Omega t)\\
= & \,\omega_{\vb*{k}, 0}\, \delta_M\, f_{cos}^{sm}(t = 0)\left(\Delta_{Re} \cos (\Omega t)+\Delta_{Im} \sin (\Omega t) \right)\\
= & \, \delta_M\, f_{\vb*{K}}\left( \Re( \omega_{\vb*{k}, 0} \Delta_{Re} ) \cos (\Omega t) + \Re( \omega_{\vb*{k}, 0} \Delta_{Im} ) \sin (\Omega t) \right) \\
 &+ i \delta_M\, f_{\vb*{K}}\left( \Im( \omega_{\vb*{k}, 0} \Delta_{Re} ) \cos (\Omega t) + \Im( \omega_{\vb*{k}, 0} \Delta_{Im} ) \sin (\Omega t) \right) \\
= &\, \delta_M\, f_{\vb*{K}} \sqrt{ \Re( \omega_{\vb*{k}, 0} \Delta_{Re} )^2 +  \Re( \omega_{\vb*{k}, 0} \Delta_{Im} )^2}\, \cos (\Omega t - \Phi_{\mathrm{Re}}) \\
 &+ i\, \delta_M\, f_{\vb*{K}} \sqrt{ \Im( \omega_{\vb*{k}, 0} \Delta_{Re} )^2 +  \Im( \omega_{\vb*{k}, 0} \Delta_{Im} )^2}\, \cos (\Omega t - \Phi_{\mathrm{Im}}).
\end{aligned}
\end{equation}
In the above we have introduced the notation $f_{\vb*{K}} \equiv f_{cos}^{sm}(t = 0)$ to highlight the strong dependence of that factor upon the mechanical Bloch wavevector. The phases $\Phi_{\mathrm{Re}}$ and $\Phi_{\mathrm{Im}}$ are given by
\begin{equation}
\begin{aligned}
\Phi_{\mathrm{Re}} & = \atan \frac{-  \Re( \omega_{\vb*{k}, 0} \Delta_{Im} ) }{ \Re( \omega_{\vb*{k}, 0} \Delta_{Re} ) }\\
\Phi_{\mathrm{Im}} & = \atan \frac{-  \Im( \omega_{\vb*{k}, 0} \Delta_{Im} ) }{ \Im( \omega_{\vb*{k}, 0} \Delta_{Re} ) }
\end{aligned}
\end{equation}

Finally, Suppl.~Eq.~(\ref{eq:QNM_temporal}) can be recast as
\begin{equation}\label{eq:QNM_final}
\begin{aligned}
\omega_{\vb*{k}} (t) = &\, \omega_{\vb*{k}, 0} + \Delta\omega(t) = \omega_{\vb*{k}, 0} \left( 1 + \frac{\Delta\omega}{\omega_{\vb*{k}, 0}}\right)\\
 = &\, {\omega_{\vb*{k}, 0}} \left( 1 + g_{\mathrm{Re}}\, \delta_M\, \cos (\Omega t - \Phi_{\mathrm{Re}}) + i\, g_{\mathrm{Im}}\, \delta_M\, \cos (\Omega t - \Phi_{\mathrm{Im}}) \right)
\end{aligned}
\end{equation}
that is the form given in the main text. 

\section{Symmetry analysis of the optomechanical coupling}
The integrals in Suppl.~Eq.~(\ref{eq:DeltaMBRe}) and Suppl.~Eq.~(\ref{eq:DeltaPERe}) contain photonic and mechanical eigenfunctions of the respective partial differential equations; as such, their symmetries are dictated by the system symmetries. Consider the system with $\beta = 0$: the unit cell has three mirror planes defined by the operations $T_{\sigma_x}$, $T_{\sigma_y}$ and $T_{\sigma_z}$. Here, $T$ are the orthogonal matrices that describe the coordinate transformation corresponding to the mirror operation. Because of the geometrical symmetry, the dielectric tensor isotropy, and limiting to eigenfunctions at the Brillouin zone center, the Maxwell equation for Bloch functions commutes with the mirror symmetry operators. Such operators are defined as $\hat{\sigma}_{\alpha} \left[ \vet E \right] (\vet r ) \equiv T_{\sigma_{\alpha}} \left[ \vet E (  T_{\sigma_{\alpha}}^{-1} \vet r ) \right] $. Commutation implies that the electric field Bloch mode is an eigenfunction of the symmetry operator: $\hat{\sigma}_{\alpha} \left[ \vet E \right] = \sigma^{E}_{\alpha}\, \vet E $ with $\sigma^{E}_{\alpha} = \pm 1$. An analogous derivation can be applied to the continuum mechanics equation, after having noticed that the unit cell geometrical symmetries are compatible with the stiffness tensor symmetries. Hence, the mechanical displacement function also obeys $\hat{\sigma}_{\alpha} \left[ \vet u \right] = \sigma^{M}_{\alpha}\, \vet u $. Concerning the stress tensor function, its nature of rank-two tensor calls for a different definition of the symmetry operator, that reads $\tilde{\sigma}_{\alpha} \left[ S_{i j} \right] (\vet r ) \equiv   T_{\sigma_{\alpha}} |_{ii'}\, T_{\sigma_{\alpha}} |_{jj'}\,  \left[ S_{i'j'} (  T_{\sigma_{\alpha}}^{-1} \vet r ) \right] $ (Einstein summation convention). With this definition, the strain tensor field is eigenfunction with the same mechanical eigenvalue: $\tilde{\sigma}_{\alpha} \left[ S_{i j} \right] = \sigma^{M}_{\alpha}\, S_{i j} $. \\
Let's start exploiting the symmetries for the evaluation of the moving boundary integral. Dividing the domain $\partial \mathcal{D}_{\mathrm{uc}}$ in two halves symmetric with respect to the $T_{\sigma_x}$ operation, the $\vet E$-term of Suppl.~Eq.~(\ref{eq:DeltaMBRe}) is written 

\begin{equation}\label{eq:symm_MB}
\begin{aligned}
& \int_{\partial  \mathcal{D}_{\mathrm{uc}}}  \ud S \Re\left[ \vet n(\vet r) \cdot \vet u (\vet r) \right]\,\vet E_{\parallel} (\vet r) \cdot \vet E_{\parallel} (\vet r)\ \Delta \varepsilon (\vet r) = \\ 
= & \int_{\partial  \mathcal{D}^{x-}_{\mathrm{uc}}}  \ud S \Re\left[ \vet n(\vet r) \cdot \vet u (\vet r) \right]\,\vet E_{\parallel} (\vet r) \cdot \vet E_{\parallel} (\vet r)\ \Delta\varepsilon (\vet r) + \int_{\partial  \mathcal{D}^{x+}_{\mathrm{uc}}} \ud S \Re\left[ \vet n(\vet r) \cdot \vet u (\vet r) \right]\,\vet E_{\parallel} (\vet r) \cdot \vet E_{\parallel} (\vet r)\ \Delta \varepsilon (\vet r) = \\
= & \int_{\partial  \mathcal{D}^{x+}_{\mathrm{uc}}}  \ud S\, |J T_{\sigma_x}| \Re\left[ \vet n(T_{\sigma_x}^{-1} \vet  r) \cdot \vet u (T_{\sigma_x}^{-1} \vet  r) \right]\,\vet E_{\parallel} (T_{\sigma_x}^{-1} \vet  r) \cdot \vet E_{\parallel} (T_{\sigma_x}^{-1} \vet  r)\ \Delta\varepsilon (T_{\sigma_x}^{-1} \vet  r)  + \int_{\partial  \mathcal{D}^{x+}_{\mathrm{uc}}} \ldots = \\
= & \int_{\partial  \mathcal{D}^{x+}_{\mathrm{uc}}}  \ud S\, \Re\left[ T_{\sigma_x} \vet n(T_{\sigma_x}^{-1} \vet  r) \cdot T_{\sigma_x} \vet u (T_{\sigma_x}^{-1} \vet  r) \right]\,T_{\sigma_x} \vet E_{\parallel} (T_{\sigma_x}^{-1} \vet  r) \cdot T_{\sigma_x} \vet E_{\parallel} (T_{\sigma_x}^{-1} \vet  r)\ \Delta\varepsilon (\vet  r)  + \int_{\partial  \mathcal{D}^{x+}_{\mathrm{uc}}} \ldots = \\
= & \left( \sigma^M_x + 1 \right)  \int_{\partial \mathcal{D}^{x+}_{\mathrm{uc}}} \ud S \Re\left[ \vet n(\vet r) \cdot \vet u (\vet r) \right]\,\vet E_{\parallel} (\vet r) \cdot \vet E_{\parallel} (\vet r)\ \Delta \varepsilon (\vet r).
\end{aligned}
\end{equation}
In this equation chain we have exploited some facts: (i) $T_{\sigma_x}$ has unit Jacobian; (ii) the dot product between two vectors is invariant upon application of $T_{\sigma_x}$ on both of them ($T_{\sigma_x}$ is an orthogonal matrix), (iii) the geometry is symmetric, which implies $\Delta\varepsilon (\vet  r) = \Delta\varepsilon (T^{-1}_{\sigma_x} \vet r)$, $ \vet n (\vet r) = T_{\sigma_x} \vet n (T^{-1}_{\sigma_x} \vet r)$, and (iv) $(\sigma^E_x)^2 = 1$. Repeated application of the above procedure also for the two other mirror planes, one gets that Suppl.~Eq.~(\ref{eq:symm_MB}) is further equivalent to
\begin{equation}
\left( 1 + \sigma^M_x \right) \left( 1 + \sigma^M_y \right) \left( 1 + \sigma^M_z \right)  \int_{\partial \mathcal{D}^{x+,y+,z+}_{\mathrm{uc}}} \ud S \Re\left[ \vet n(\vet r) \cdot \vet u (\vet r) \right]\,\vet E_{\parallel} (\vet r) \cdot \vet E_{\parallel} (\vet r)\ \Delta \varepsilon (\vet r).
\end{equation}
Since an analogous treatment can be done on the term containing the $\vet D$ function, one finally gets Eq.~(6) of the main text. There, we have labeled $\partial \mathcal{D}^{x+,y+,z+}_{\mathrm{uc}}$ as $\partial \mathcal{D}_{1/8}$ for notation brevity. 

When dealing with the photoelastic integral, a treatment similar to that given above can be made, except for the step where the invariance of dot product is exploited. Now, one has to use more explicitly the orthogonality of matrices $T_{\sigma_{\alpha}}$ (from now on written $T$ for brevity). Using $T_{ii'}T_{ii''} = \delta_{ii''}$ (Einstein summation convention), one has
\begin{equation}
\begin{aligned}
& \int \ud V\, S_{i j}(T^{-1} \vet r)\,  E_{k} (T^{-1} \vet r)\, E_{l} (T^{-1} \vet r)\, p_{ijlk}(T^{-1} \vet r) = \\
= &\int \ud V\, T_{i i'} T_{j j'} S_{i j}(T^{-1} \vet r)\, T_{k k'} E_{k} (T^{-1} \vet r)\, T_{l l'} E_{l} (T^{-1} \vet r)\, T_{i i''} T_{j j''} T_{k k''} T_{l l''}\ p_{i''j''l''k''}(T^{-1} \vet r) .
\end{aligned}
\end{equation}
Now, one has to ascertain that $T_{i i''} T_{j j''} T_{k k''} T_{l l''}\ p_{i''j''l''k''}(T^{-1} \vet r) = p_{ijkl}( \vet r) $, which is indeed  true as it was a prerequisite to have the operator $\tilde{\sigma}$ commuting with the continuum mechanics equation. In substance, also the photoelastic term is given by an integral over one eighth of the unit cell, with the prefactor $\left( 1 + \sigma^M_x \right) \left( 1 + \sigma^M_y \right) \left( 1 + \sigma^M_z \right)$. 

In the case of finite slant angle, $\beta \neq 0$, only the factors $\left( 1 + \sigma^M_x \right) \left( 1 + \sigma^M_y \right)$ remain, thus leading to the behavior described in main Fig.~4.

\subsection{Justification of the single-mode perturbation approach}
We have earlier worked within the hypothesis that a single-mode perturbation theory was valid; we here justify that by means of symmetry analysis, at least for the modes of our concern. When multiple electromagnetic modes are present in the frequency range of interest, Suppl.~Eq.~(\ref{eq:QNM_PT}) is not valid and one should instead diagonalize a perturbative matrix containing off-diagonal terms like $\int \vb*{E}^L_i\, \Delta\varepsilon\, \vb*{E}^R_j\, \mathrm{d}V $. In our case, at zone center there are two quasi-degenerate modes (Suppl.~Sect.~\ref{sect:characterization}), whose symmetry eigenvalues are
\begin{center}
\begin{tabular}{cc|cc}
Mode index & Frequency & \ $\sigma^E_x$ & $\sigma^E_y$ \\
\hline
\textbf{1} & 194.53 THz\ & - & + \\
\textbf{2} & 194.59 THz\ & + & - \\
\hline 
\end{tabular}
\end{center}
Applying the previously described procedure, one gets 
\[
\int_{\mathcal{D}_\mathrm{uc}} \vb*{E}^L_1\, \Delta\varepsilon\, \vb*{E}^R_2\, \mathrm{d}V = \left( 1 + \sigma^M_x \sigma^E_{x,1} \sigma^E_{x,2} \right) \left( 1 + \sigma^M_y \sigma^E_{y,1} \sigma^E_{y,2} \right) \int_{\mathcal{D}_\mathrm{uc}^{x+,y+}} \vb*{E}^L_1\, \Delta\varepsilon\, \vb*{E}^R_2\, \mathrm{d}V 
\]
where $\sigma^E_{\alpha,1}$ and $\sigma^E_{\alpha,2}$ refer to the symmetry eigenvalue of operation $\hat{\sigma}_{\alpha}$ on the electric fields of modes 1 and 2. 
Given the table above, the integral is zero if the mechanical mode does not have simultaneously $\sigma^M_x=-1$ and $\sigma^M_y=-1$. Hence, single-mode perturbation theory is justified for the majority of mechanical modes listed in main Figs.~3 and 4, exception made in the vicinity of mechanical modes number 8, 11, and 15. We would highlight that the present analysis is valid if electromagnetic and mechanical modes are close to the $\Gamma$ point; further work is required to analyze the general configuration. We however expect a sufficiently smooth behavior as $\vet K$ and $\vet k$ are moved away from the $\Gamma$ point, thus justifying the numerical observations, especially concerning mechanical mode 7.

\begin{figure*}
\centering
\includegraphics[width = \textwidth]{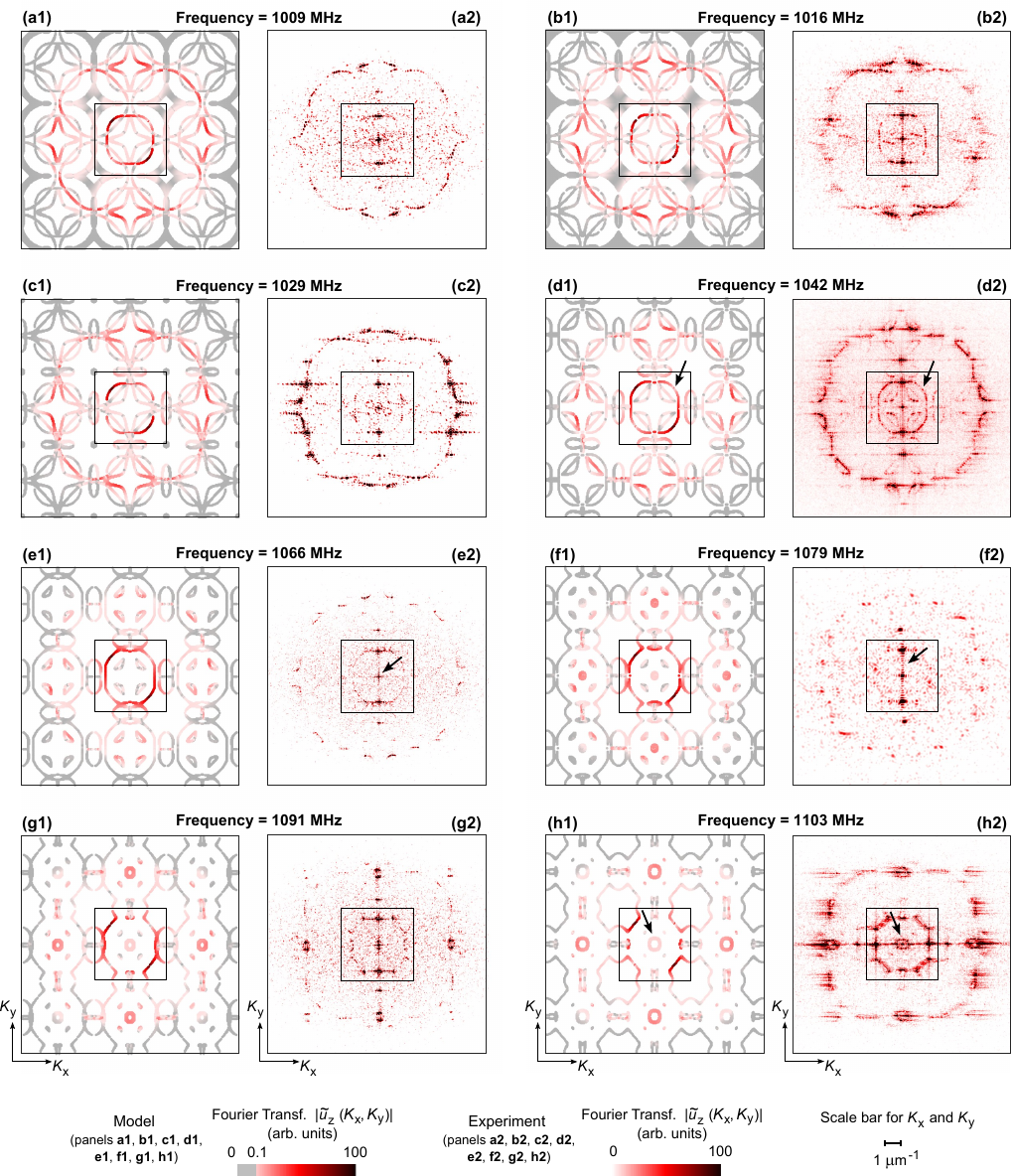}
\caption{Reciprocal space maps showing the population of mechanical Bloch modes.}\label{fig:fouriermaps}
\end{figure*}

\section{Numerical FEM calculations}
The Bloch mode dispersion and optomechanical couplings have been calculated within a FEM model implemented under the COMSOL environment. Our scripts are partially based on those available with \cite{Primo2020}. The geometry consists of an unit cell of the metasurface ($\mathcal{D}'_{\mathrm{uc}}$ in Suppl.~Fig.~\ref{fig:domains}). Maxwell equations are solved in the whole $\mathcal{D}'_{\mathrm{uc}}$, while the continuum mechanics equation is solved in $\mathcal{D}_{\mathrm{uc}}$. The RF module and Solid Mechanics module have been employed, respectively. In the RF module, GaAs is modeled as a dielectric with nondispersive relative permittivity, $\varepsilon_{\mathrm{GaAs}} = 11.44$. In the Solid Mechanics module, GaAs is modeled as a Piezoelectric material with elasticity matrix given by $c_{11} =c_{22} =c_{33} = 11.88 \cdot 10^{10}\, \mathrm{Pa}$, $c_{12} =c_{13} =c_{23} = 5.38 \cdot 10^{10}\, \mathrm{Pa}$, $c_{44} =c_{55} =c_{66} = 5.94 \cdot 10^{10}\, \mathrm{Pa}$ (see Ref.~\cite{YuCardona}). The coupling matrix is $e_{14} =e_{25} =e_{36} = 0.154\, \mathrm{C/m}^2$ (stress-charge form used). Those values are stated with reference to the conventional GaAs cubic cell. GaAs mass density is $5320\, \mathrm{kg/m}^2$. The mesh is triangular with maximum element size of $70\, \nm$ in the GaAs domain and $200\, \nm$ elsewhere.

\section{Mechanical Bloch mode population analysis}
To obtain better understanding of the optomechanical coupling, we performed a purely mechanical characterization of the displacement field excited in the metasurfaces. The method employed is based on interferometry; for details see for instance \cite{ZanottoNatComm2022}. In essence, real-space mapping of the out-of-plane displacement component is performed, obtaining maps $\tilde{u}_z(x,y,z^{*},t^{*})$ where $z^{*}$ is the membrane surface and $t^{*}$ is a fixed reference time in the mechanical cycle. Fourier transforming those maps leads to reciprocal space maps $\tilde{u}_z(K_x,K_y)$, that are plotted in Suppl.~Figs.~\ref{fig:fouriermaps} a1-h1. Those maps should be compared with Suppl.~Figs.~\ref{fig:fouriermaps} a2-h2, obtained from numerical modeling. An important feature of those maps is that the theoretical ones indicate the modes that \textit{are in principle allowed} in the crystal, while the experimental ones indicate the modes that \textit{are actually excited}. Moreover, the theoretical maps are obtained assuming that all the Bloch modes are populated with unity amplitude, while the experimental ones clearly reflect the actual population of each Bloch mode. Significantly, in the experiment, Bloch mode population is observed for almost all the directions in the reciprocal space, despite the wave excited by the IDTs is directional ($K_x = 0$). We have already noticed this effect in very similar samples \cite{ZanottoNatComm2022}, and attributed the effect to (i) the non perfectly planar interface between bulk GaAs (where the IDT is located) and the metasurface membrane, and (ii) the scattering inside the metasurface originating from fabrication-induced inhomogeneities. The data in Suppl.~Fig.~\ref{fig:fouriermaps} support the analysis presented about Fig.~3 of the main text. Indeed, the excitation of mode 3 is present at all frequencies (that mode is the ring-shaped structure evidenced in Suppl.~Fig.~\ref{fig:fouriermaps}d1-2, evolving in more complex shapes at higher frequencies). Moreover, mode 7 is clearly visible in Suppl.~Figs.~\ref{fig:fouriermaps}h1-2 (small ring centered at the origin). Finally, the presence of finite experimental signal at $(K_x,K_y)$ locations where theory does not predict Bloch modes (see for instance the arrows in Suppl.~Figs.~\ref{fig:fouriermaps}e2, \ref{fig:fouriermaps}f2) suggests the hypothesis that, in addition to the Bloch modes of a perfectly periodic structure, other modes such as disorder-induced modes are present.

\section{Effect of finite slant angle on the mechanical band structure}
\begin{figure*}
\centering
\includegraphics{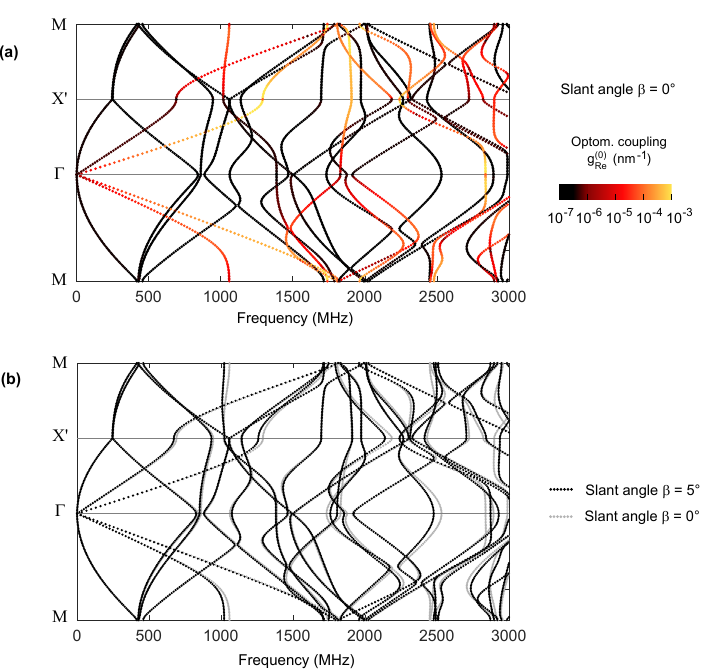}
\caption{(a), Mechanical mode dispersion and optomechanical coupling for the structure with zero slant angle. (b), Comparison between the mechanical mode dispersion for slanted and non-slanted structures.}\label{fig:mechbands}
\end{figure*}

We report here (Suppl.~Fig.~\ref{fig:mechbands}a) the analogous of main Fig.~3d, calculated with zero slant angle $\beta$. 
Consistently with main Fig.~4a, most of the modes have zero coupling. Interestingly, this happens not only at the $\Gamma$ point but throughout the whole Brillouin zone. We also report (Suppl.~Fig.~\ref{fig:mechbands}b) a direct comparison between the mechanical mode dispersion of the structures with slanted and non-slanted holes. Similarly to what observed for photonic modes (Suppl.~Fig.~\ref{fig_photonicbands}), the slant does not affect the global mode structure. 

\section{Spatial mapping of the optomechanical modulation}
\label{sect:SpatialMapping}
\begin{figure*}[h]
\centering
\includegraphics{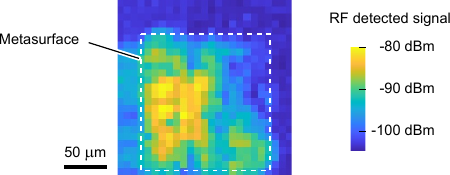}
\caption{Spatially-resolved mapping of the transmission modulation performed with a focused laser beam.}\label{fig:transdmap}
\end{figure*}

To have further insights into the role of inhomogeneities and scattering on the transmittance modulation, we have performed a spatially resolved measurement of the optomechanical signal. To improve the spatial resolution (at the expense of a lower angular resolution) we employed a higher numerical aperture, more tightly focused laser beam ($\approx 20\ \um$). To this purpose we replaced the 50 mm focusing lens (see Suppl.~Sect.~\ref{sect:ExpSetup}) with a shorter focal one (30 mm achromatic doublet) and collected the map illustrated in Suppl.~Fig.~\ref{fig:transdmap}. An incident angle of $6^{\circ}$ was employed, and a RF frequency of 1102.75 MHz. Some transduction hotspots can be observed, few of them having a size of the order of $20\ \um$. This spatial structure supports the effectively coupled region picture (see Suppl.~Sect.~\ref{sect:QNMPT}), with a typical size $\sigma^{\mathrm{eff}}$ definitely smaller than the metasurface itself, and also smaller than the $30\ \um$ laser spot size employed for the other modulation measurements.  We would highlight that the size of the features in Suppl.~Fig.~\ref{fig:transdmap} is not, however, a direct characterization of the effectively coupled region size $\sigma^{\mathrm{eff}}$, as a complete analysis would involve the study of interference between the multiply populated Bloch modes and the structure disorder. We nonetheless believe that the observed $20\ \um$ feature size safely sets an upper limit for $\sigma^{\mathrm{eff}}$.

\section{Estimate of the mechanical displacement within the metasurface}
\begin{figure*}[h]
\centering
\includegraphics{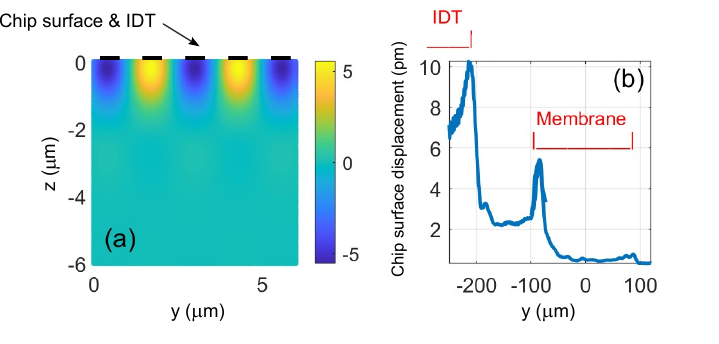}
\caption{Panel (a), simulation of the displacement field below the IDTs (color map, range $\pm 5.2\ \mathrm{pm}$). Panel (b), measurement of the displacement field over a path extending from the IDT to the metasurface membrane. }\label{fig:displestimate}
\end{figure*}

The mechanical displacement within the metasurface has been estimated combining surface acoustic wave transfer matrix simulations and interferometer-based measurements. Considering an electrical RF bias of 20 mV (i.e.~-21 dBm RF power) and a S-parameter for reflection ($S_{11}$) of -0.5 dB at resonance, we can estimate a total power for the generated SAW of $3.56\ \mu \mathrm{W}$, which gives a linear power density of $0.01\ \mathrm{W/m}$ for our $180\ \micron$ wide IDTs. Considering this power density, simulation returns a SAW with a peak-to-peak amplitude of $5.94\ \mathrm{pm}$, with negligible changes in the frequency range in our experiment. The simulated SAW vertical displacement is plotted in Suppl.~Fig.~\ref{fig:displestimate}(a). The simulated amplitude is the one generated at one of the IDT’s edges; the actual amplitude on sample will depend on wave propagation from the IDT to the patterned region. Roughly, this can be estimated by considering typical interferometry measurements such as the one reported in Suppl.~Fig.~\ref{fig:displestimate}(b). Here the SAW vertical displacement (in arbitrary units in the experiment) has been rescaled to obtain the simulated value at the end of the IDT. After a kink in the signal at the interface between bulk and suspended membrane, the signal stabilizes to an average value of $0.28\ \mathrm{pm}$. This represents an average estimate, which can be affected by the local metasurface pattern and by local device imperfections. 

